\documentclass[letterpaper,12pt,titlepage,oneside,final]{book}

\def\onedot{$\mathsurround0pt\ldotp$}
\def\cdddot#1{
  \mathbin{\vcenter{\baselineskip.67ex
    \hbox{\onedot}\hbox{\onedot}\hbox{\onedot}%
  }}%
}
 
\newcommand{\href}[1]{#1} 

\usepackage{ifthen}
\newboolean{PrintVersion}
\setboolean{PrintVersion}{false}

\usepackage{amsmath,amssymb,amstext} 
\usepackage[pdftex]{graphicx}
\usepackage{array}
\usepackage{xfrac}
\usepackage[thinlines]{easytable}
\usepackage{float}
\usepackage{multirow}
\usepackage{array}
\usepackage{breqn}
\usepackage{bm}
\usepackage[numbers, square, sort&compress]{natbib}
\usepackage[utf8]{inputenc}
\usepackage[T1]{fontenc}
\usepackage{lmodern}
\usepackage{graphicx}
\usepackage{color}
\usepackage[pdftex,pagebackref=false]{hyperref} 
\usepackage{amsmath}
\usepackage{amsfonts}
\usepackage{epstopdf}
\usepackage{matlab}
\usepackage{blindtext} 
\usepackage[font=footnotesize,labelfont=bf]{caption}
\usepackage{subcaption}
\usepackage{enumitem}
\usepackage{siunitx}
\usepackage{booktabs}
\usepackage{colortbl} 
\usepackage{makecell}
\usepackage{float}
\usepackage{hhline}

\definecolor{mydarkgray}{RGB}{150,150,150} 
\definecolor{mywhite}{RGB}{255,255,255}
\definecolor{mylightgray}{RGB}{220,220,220}
\definecolor{mylightblue}{RGB}{173,216,230} 
\definecolor{mylightpink}{RGB}{255,224,230} 
\definecolor{mylightyellow}{RGB}{255,255,204} 
\definecolor{mylightgreen}{RGB}{216,237,216} 
\definecolor{mymediumgray}{RGB}{211,211,211} 

\newcolumntype{P}[1]{>{\centering\arraybackslash}p{#1}}

\hypersetup{
    plainpages=false,       
    unicode=false,          
    pdftoolbar=true,        
    pdfmenubar=true,        
    pdffitwindow=false,     
    pdfstartview={FitH},    
    pdfnewwindow=true,      
    colorlinks=true,        
    linkcolor=blue,         
    citecolor=green,        
    filecolor=magenta,      
    urlcolor=cyan           
}
\ifthenelse{\boolean{PrintVersion}}{   
\hypersetup{	
    citecolor=black,%
    filecolor=black,%
    linkcolor=black,%
    urlcolor=black}
}{} 

\usepackage[abbreviations]{glossaries-extra} 

\let\origdoublepage\cleardoublepage
\newcommand{\clearemptydoublepage}{%
  \clearpage{\pagestyle{empty}\origdoublepage}}
\let\cleardoublepage\clearemptydoublepage

\begin{document}

\pagestyle{empty}
\pagenumbering{roman}

\begin{titlepage}
    \begin{center}
        \LARGE\textbf{High Accuracy Determination of Rheological Properties of Drilling Fluids Using the Marsh Funnel}

        \vspace{0.5cm}

        \large
        Sanket Biswas\footnote{Email: \texttt{1940108@sliet.ac.in}, \texttt{sanketb@mail.ubc.ca}}, 
        Harshita Tiwari\footnote{Email: \texttt{1940097@sliet.ac.in}, \texttt{ch23s012@smail.iitm.ac.in}}, Sujata Verma\footnote{Email: \texttt{1940105@sliet.ac.in}}, and
        Kamlesh Kumari\footnote{Email: \texttt{kamlesh213@sliet.ac.in}}
        
        \vspace{0.2cm}
        
        \small
        \textsuperscript{\textcolor{blue}{1, 2, 3, 4}} \textit{Department of Chemical Engineering, Sant Longowal Institute of Engineering and Technology, Longowal, Sangrur, PB 148106, India}
        
        \textsuperscript{\textcolor{blue}{1}} \textit{Department of Chemical and Biological Engineering, University of British Columbia, 2360 E Mall, Vancouver, BC V6T 1Z3, Canada}
        
        \textsuperscript{\textcolor{blue}{2}} \textit{Department of Chemical Engineering, Indian Institute of Technology Madras, Chennai, TN 600036, India}
        
        \textsuperscript{\textcolor{blue}{3}} \textit{Khanna Paper Mills Limited, NH 3, Kamla Devi Avenue, Amritsar, PB 143001, India}

        \vspace{0.2cm}
        \footnotesize
        \begin{minipage}{\textwidth}
        \textbf{Abstract.} Efficient and safe drilling operations require precise determination of rheological properties in drilling fluids, encompassing dynamic viscosity for Newtonian fluids, and apparent viscosity, plastic viscosity, and yield point for non-Newtonian fluids. Conventional viscometers like vibrating wire, ZNN-D6, and Fann-35 offer high accuracy but are limited by cost and complexity in small-scale industries and labs. To address this, our research presents a novel mathematical model based on the Herschel-Bulkley model, aiming to accurately characterise drilling fluids' rheological properties using the Marsh funnel as an alternative device --- an economical, operator-friendly, and power-independent equipment. Drawing inspiration from seminal works by \cite{li2020rheological}, \cite{sedaghat2017novel}, and \cite{guria2013rheological}, this innovative framework establishes a universal inverse linear relationship between a fluid's flow factor and final discharge time. For any fluid, it utilises its density and flow factor (or final discharge time) to determine all its rheological properties. Specifically, it evaluates dynamic viscosity for Newtonian fluids, apparent viscosity, plastic viscosity, and yield point for weighted non-Newtonian fluids, and apparent viscosity for non-weighted non-Newtonian fluids, with average systematic errors (against Fann-35 measurements) of 0.39\%, 3.52\%, 2.17\%, 18.38\%, and 5.84\%, respectively, surpassing the precision of alternative mathematical models found in the aforementioned literature. Furthermore, while our framework's precision in plastic viscosity and yield point assessment of non-weighted non-Newtonian fluids slightly lags behind the framework of \cite{li2020rheological}, it outperforms the model of \cite{sedaghat2017novel}. In conclusion, despite minor limitations, our proposed mathematical model holds huge promise for drilling fluid rheology in petroleum, drilling, and related industries.
        \end{minipage}

        \vspace{0.2cm}
        \begin{minipage}{\textwidth}
        \end{minipage}
    \end{center}
\end{titlepage}

\pagestyle{plain}
\setcounter{page}{2}

\cleardoublepage

  
  

  


 \cleardoublepage


\Large \begin{center}\textbf{Acknowledgements}\end{center}
\normalsize
The authors express their sincere gratitude to the Indian Institute of Chemical Engineers (IIChE) for their benevolent provision of the IIChE Research Grant, which served as a critical enabler for the successful realisation of this research initiative. Additionally, the authors extend their appreciation to the Department of Chemical Engineering at Sant Longowal Institute of Engineering and Technology (SLIET) for their gracious facilitation of access to their advanced laboratory facilities and state-of-the-art equipment throughout the entire duration of this research endeavour.

Furthermore, the authors acknowledge with thanks the invaluable cooperation of SD Fine Chem Limited (SDFCL) and Loba Chemie Private Limited, whose collaboration enabled the acquisition of essential experimental chemicals at markedly subsidised rates, significantly contributing to the experimental progress and outcomes of this study.

\cleardoublepage

  


 \cleardoublepage

\renewcommand\contentsname{Table of Contents}
\tableofcontents
\cleardoublepage
\phantomsection    

\pagenumbering{arabic}

\chapter{Introduction}
\label{chap:1}
\section{Motivation}
Drilling fluids are necessary at practically every stage in the drilling process, or, more generically, in the oil field. Drilling efficiency and safety are significantly impacted by the rheological characteristics of drilling fluids, which essentially comprise \textit{apparent viscosity}, $\mu_a$, \textit{plastic viscosity}, $\mu_p$, and \textit{yield point}, $\tau_0$. So, for effective drilling operations, mud engineers must frequently evaluate and track the aforementioned properties of drilling fluids using either capillary, vibrating wire, or rotational viscometers (which include the well-known ZNN-D6 and Fann-35 viscometers), or Marsh funnels \cite{marsh1931properties}.

Capillary viscometers can operate only under very specialised experimental settings, and rotational and vibrating wire viscometers have complicated structures, require a power supply for measurements, and are expensive. In field operations, the usage of the aforementioned viscometers is constrained by these factors, particularly for small-scale drilling rigs. Marsh funnels, in contrast, are less expensive, do not require sophisticated testing apparatus, and require little laborious lab work. Due to these benefits, Marsh funnels are frequently used to assess the rheological characteristics of drilling fluids and express the fluid's quality in terms of its \textit{Marsh funnel viscosity (MFV)}. However, it is not a true viscosity and is defined as the amount of time (in seconds) needed for one quart of drilling mud to flow out of a Marsh funnel into a graded mud cup. As a result, it is solely utilised for qualitative analysis and lacks any inherent relevance, and little has been done to determine the above rheological properties using it, until the last two decades or so. In light of this, building upon the seminal works of \cite{balhoff2011rheological}, \cite{guria2013rheological}, \cite{sedaghat2017novel} and \cite{li2020rheological}, our main motive is to expand the function of the Marsh funnel, given its simplicity, to reliably quantify the aforementioned rheological properties of most, if not all, Newtonian and non-Newtonian drilling fluids. In the next section, we briefly review all the previous relevant research works.

\section{Background}
With the exception of a few well-known Newtonian fluids, such as water, air, and some oil products, most fluids used in the drilling industry are non-Newtonian. To date, there have been many rheological models put forth to describe the flow properties of non-Newtonian fluids using the Marsh funnel. To begin with, in 2000, \cite{pitt2000marsh} proposed a correlation between the measured MFV ($t$ in s), density ($\rho$ in g/cm$^3$) and effective viscosity ($\mu_e$ in cP) of non-Newtonian power-law fluids, given by 
\begin{equation}
    \mu_e = \rho(t-25).
    \label{mue}
\end{equation} \cite{almahdawi2014apparent} extended this work by developing a novel model that determines the rheological characteristics of drilling fluids and other non-Newtonian fluids using fluid density and MFV. They established the following empirical correlation between apparent viscosity (in cP), fluid density (in g/cm$^3$) and MFV (in s): \begin{equation}
    \mu_a = \rho(t-28).
    \label{mua}
\end{equation}
The rheological properties of polymer melts and solutions, a sizable group of very viscous liquids with non-Newtonian and viscoelastic natures, were intensively researched by \cite{brydson1970flow} and \cite{carreau2021rheology}. \cite{barnes1985yield} and \cite{barnes1989introduction} further addressed the rheology of complex non-Newtonian fluids such as polymeric liquids and suspensions. \cite{barnes1989shear}, \cite{barnes1997thixotropy}, \cite{barnes1999yield} investigated the yield stress of non-Newtonian fluids, reviewed thixotropy, and analysed shear-thickening (dilatancy) in suspensions. 

The long-standing issue of interpreting the rheological characteristics of polymer support fluids from Fann-35 viscometer test results was addressed in \cite{lam2014interpretation}. They introduced two conversion factors that can be used to present viscometer data in SI units: the \textit{flow consistency index} and the \textit{flow behaviour index}. The flow indices were coupled to the Fann-35 viscometer readings at $\theta_{300}$ and $\theta_{600}$ at rotational speeds of $300$ rpm and $600$ rpm, respectively, after analysis of the shear stress and shear rate curves.

A comprehensive methodology for measuring the rheological characteristics of cement pastes and grouts, a class of Bingham fluids, was published in \cite{roussel2005marsh}. They showed that the \textit{discharge time}, $t$, can be directly linked to the material behaviour, including the above rheological properties, and is based on a technique using two simple funnels differing by their nozzles. Through this study, they demonstrated that this methodology can be widely used in fields for quality control of cement pastes and grouts, and other industrial Bingham fluids.

A finite element method was used to predict the flow of new cement suspensions inside the Marsh funnel in the study by \cite{cremonesi2010simulation}. They captured images of the funnel's flow at various discharge times. They noted that it was intriguing to observe that the fluid's free surface in the Marsh funnel was not perfectly flat. Additionally, they identified some unyielded regions
within the fluid which showed the three-dimensionality of fluid flow in the Marsh funnel for certain (non-Newtonian) suspensions.

\cite{balhoff2011rheological} presented a new method for obtaining rheological properties of drilling muds and other non-Newtonian fluids that employs height versus
time data in a draining Marsh Funnel, acquired by applying mass balance around the fluid control volume. In combination with the expression for pressure drop, they formulated an explicit expression of the average wall shear stress, $\langle\tau_w\rangle$. The ODE obtained for Newtonian fluids (mineral oil in this case), was solved using the Hagen-Poiseuille equation; for non-Newtonian fluids, such as xanthan polymer, bentonite muds and hydroxy polyacrylamide (HPAM), Runge-Kutta method was applied to solve the ODE numerically and obtain the rheological properties using the measured height versus time data. They further plotted the liquid height versus discharge time data of different Newtonian and non-Newtonian fluids and used fitting functions (mostly least-squares fit) to fit the height-time variation. Moreover, they compared these with the funnel and prediction data for validation. They obtained the viscosity ($\mu_a$ and $\mu_p$) of shear-thinning fluids in excellent agreement with Fann-35 viscometer measurements. Furthermore, they used the \textit{ultimate static height} in the Marsh funnel to calculate the yield stress, $\tau_0$. However, in this research, the average wall shear stress values deviate significantly from the actual data primarily due to the exclusion of the height of the tube section from the fluid height, $h$ (assuming the height of the tube section, $H_T$, is negligible as compared to the height of the conical section, $H_C$).

\cite{guria2013rheological} developed a method to estimate the average shear stress, $\langle\tau_w\rangle$, and the average shear rate, $\langle\dot{\gamma}_w\rangle$, on the walls of the Marsh funnel from the fluid discharge volume and rate, respectively. The discharge volume and the corresponding discharge time were the only two measured variables used in their study. Assuming no-flow conditions, they computed the yield point, $\tau_0$, from the liquid height. Further, they determined the apparent and plastic viscosity from the funnel consistency plots. To ascertain the aforementioned rheological characteristics, they studied synthetic crude oil and a number of drilling fluid additives, including bentonite, PEG-NaCl, and PEG-NaCl-bentonite, and they compared the findings with the Fann-35 viscometer readings. Even though the percentage systematic error in the above comparison is considerable, this methodology has been widely adopted by many researchers due to its tremendous potential for extension and improvement.

For the analysis of the rheological characteristics of bentonite suspensions, \cite{schoesser2015marsh} employed the same methodology as \cite{guria2013rheological}. In their research, they developed a mathematical model by adopting the mathematical equations of capillary viscometry. This model was used to calculate the discharge rate (at the outlet) and yield stress as a function of fluid height, and plot shear stress vs shear rate curves (also known as \textit{characteristic flow curves}). Using these and the constitutive equations of the Herschel-Bulkley model, the authors calculated the yield point of the relevant bentonite suspensions. However, their results deviated considerably from the corresponding Fann-35 viscometer measurements.

\cite{liu2003measuring} provided a theoretical basis for calculating the rheological parameters of power-law fluids, particularly drilling fluids, using a funnel viscometer. Using this device, experiments were conducted with four types of fluids, and the relevant rheological parameters were found to be in good agreement with those of rotational viscometer measurements. Based on this model, \cite{liu2014measure} established a relationship between the change in shear stress and the change in shear rate by measuring the \textit{final discharge time}, $t_F$. Using this, they determined the apparent viscosity and plastic viscosity of power-law fluids and also extended the formulation of \cite{guria2013rheological} for determining the yield point, $\tau_0$.

From the above literature review, it is clear that there has been an increasing interest in using the Marsh funnel to determine the rheological properties of drilling fluids. However, significant discrepancies in results when compared to other standard techniques have raised concerns about the accuracy of the aforementioned models based on the Marsh funnel method. As a result, a more detailed analysis of the Marsh funnel's geometry seemed to be inevitable. Furthermore, it appeared that there is a gap in the development of a simpler model to determine the average shear stress on the walls of the Marsh funnel and the corresponding average shear rate. In an attempt to address these inaccuracies associated with the Marsh funnel method, \cite{sedaghat2016mathematical} proposed a mathematical model that precisely calculates the volume of the Marsh funnel using a third-order polynomial function. The frameworks discussed earlier had some inaccuracies due to simplified models for the cone volume, which potentially impacted the accuracy of rheological properties that are sensitive to even small variations of liquids within the Marsh funnel, including yield point and apparent and plastic viscosity. Further, they developed a flow rate model using a parameter known as the \textit{flow factor}, $f$, that is determined using the final discharge time, $t_F$. With this factor, the temporal height of the fluid in the Marsh funnel was also formulated. This seminal work represented an important step towards improving the accuracy and reliability of the Marsh funnel method for determining the rheological properties of drilling fluids and paved the way for further research and improvements in this area.

\cite{li2020rheological} expanded on the work of \cite{guria2013rheological} and \cite{sedaghat2017novel} by developing simplified mathematical models for average wall shear rate, $\langle\dot{\gamma}_w\rangle$, and average wall shear stress, $\langle\tau_w\rangle$, and proposing mathematical models for rheological parameters of Newtonian and non-Newtonian fluids based on the Herschel-Bulkley model. They also introduced a new correction factor for the wall shear rate of weighted non-Newtonian fluids. In addition, they proposed a simple correlation between the flow factor and the final discharge time:
\begin{equation}
f\cdot t_F = 38,
\label{ftf}
\end{equation}
and used it to characterize the rheological properties of seventeen selected drilling fluids using a Marsh funnel. Moreover, they implemented a CFD modelling approach and used it to establish a two-phase flow 3D model. The results showed that the proposed models calculated the dynamic, apparent, and plastic viscosities of the above fluids with an average systematic error of 1.35\%, 8.18\%, and 5.42\%, respectively, while the average systematic error of the yield stress was relatively higher (21.43\%) but within an acceptable range. This study demonstrated the potential of the proposed models to calculate the rheological parameters of drilling fluids directly and almost accurately, with only the need to measure the final discharge time and density.

\section{Highlights of the present study}
In the present study, we propose a mathematical model based on the models of \cite{li2020rheological}, \cite{sedaghat2017novel} and \cite{sedaghat2016mathematical} to determine the dynamic viscosity of Newtonian drilling fluids and apparent and plastic viscosity of non-Newtonian drilling fluids with higher accuracy than established by \cite{li2020rheological}. In addition, the present mathematical framework determines the yield point of non-Newtonian drilling fluids with almost the same accuracy as that of \cite{li2020rheological}. Similar to \cite{li2020rheological}, the flow factor and the final discharge time are correlated via a simple mathematical equation, and the above rheological properties are determined by simply measuring the fluid density and the final discharge time. In the subsequent chapter, we will provide a detailed discussion of this framework. However, before delving into the specifics of our proposed model, in the following sections, we present a comprehensive overview of the types of drilling fluids and the widely used rotational viscometer-based methods currently being employed in the field for determining the rheological properties of drilling fluids. This discussion will serve to contextualise our research and provide a foundational understanding of the various drilling fluids and their properties, as well as the current state-of-the-art in measuring these properties. This overview will provide a necessary background for understanding our proposed mathematical framework and its potential impact on the field of drilling fluid rheology.

\section{Classification of drilling fluids}
\label{sec: classification}
The classification of drilling fluids is important because it can impact their performance in terms of lubricity, rheology, and other properties that are critical for successful drilling operations. The choice of drilling fluid depends on factors such as the formation being drilled, the desired drilling speed, and the environmental regulations in place.

The classification of drilling fluids can be approached through various methods, one of which is based on their composition, properties, and functions. In light of this, let us first briefly discuss this particular categorization scheme as follows:
\begin{enumerate}
    \item \textbf{Water-based drilling fluids:} These comprise water, clays, and various chemicals. They are widely used in drilling operations due to their low cost, availability, and ease of use.

    \item \textbf{Oil-based drilling fluids:} These comprise base oils, such as diesel or mineral oils, and additives such as emulsifiers and wetting agents. They are used in drilling operations that require higher lubricity and better temperature stability.

    \item \textbf{Synthetic-based drilling fluids:} These comprise synthetic materials, such as esters, glycols, and polyalphaolefins. They are used in drilling operations where environmental concerns are paramount or where water-based drilling fluids cannot be used due to temperature or pressure limitations.

    \item \textbf{Gas-based drilling fluids:} These comprise gases, such as nitrogen or air, and additives such as foaming agents. They are used in drilling operations where wellbore stability is critical, and they help reduce formation damage.

    \item \textbf{Polymer-based drilling fluids:} These comprise polymers, such as xanthan gum or starch, and additives such as biocides and surfactants. They are used in drilling operations where shale inhibition is critical, and they help reduce drilling problems such as a stuck pipe.

    \item \textbf{Invert emulsion drilling fluids:} These comprise water droplets dispersed in an oil-based continuous phase. They are used in drilling operations where high-temperature and high-pressure conditions are present.

    \item \textbf{Acidic drilling fluids:} These comprise acids, such as hydrochloric acid or formic acid, and additives such as corrosion inhibitors. They are used in drilling operations where the formation needs to be cleaned or stimulated.

    \item \textbf{Alkaline drilling fluids:} These comprise alkaline substances, such as sodium hydroxide or potassium hydroxide, and additives such as surfactants. They are used in drilling operations where the formation is acidic, and they help reduce corrosion and improve drilling performance.
\end{enumerate}

Apart from the aforementioned classification scheme, another significant categorisation approach for drilling fluids pertains to their rheological behaviour, which is particularly pertinent to the current research. Based on this criterion, drilling fluids can be classified into two principal categories:
\begin{enumerate}
    \item \textbf{Newtonian fluids:} These fluids have a constant viscosity regardless of the shear rate applied. In other words, the viscosity of a Newtonian fluid remains the same under all flow conditions. Putting this mathematically, there is a linear relationship between the (wall) shear stress, $\tau_w$ (Pa), and the (wall) shear rate, $\dot{\gamma}_w$ (s$^{-1}$), for Newtonian (drilling) fluids, given by
    \begin{equation}
        \tau_w = \mu \dot{\gamma}_w,
        \label{Newtondrill}
    \end{equation}
    where $\mu$ (Pa$\cdot$s) is the dynamic viscosity. Examples of Newtonian fluids used in drilling include water, brine, diesel, and some oil-based drilling fluids.

    \item \textbf{Non-Newtonian fluids:} These fluids exhibit a variable viscosity in response to applied shear stress. In other words, their viscosity changes with the shear rate applied. Mathematically, this means that there is a nonlinear relationship between the (wall) shear stress, $\tau_w$, and the (wall) shear rate, $\dot{\gamma}_w$, and is given by the Herschel-Bulkley model \cite{american1997recommended}: 
    \begin{equation}
        \tau_w = \tau_0 + \eta \dot{\gamma}_w^n,
        \label{HerBul}
    \end{equation}
 where the yield point, $\tau_0$, the consistency index, $\eta$, and the flow index, $n$, describe the non-linear nature of non-Newtonian fluids. 
 
 Non-Newtonian drilling fluids can be further classified into two subcategories:
    \begin{enumerate}
        \item \textbf{Non-weighted non-Newtonian fluids:} These fluids exhibit shear-thinning behaviour, meaning that their apparent viscosity decreases as the shear rate increases. They are typically low in density and are commonly used in situations where low formation damage is desired. These fluids usually contain polymers as the primary rheology modifier and are commonly used in shallow drilling operations. Examples include xanthan gum, guar gum, hydroxyethyl cellulose (HEC), carboxymethyl cellulose (CMC), and organophilic clays.

        \item \textbf{Weighted non-Newtonian fluids:} These fluids are similar to non-weighted non-Newtonian fluids but have a higher density due to the addition of weighting agents, such as barite or hematite. These fluids exhibit shear-thickening behaviour, meaning that their apparent viscosity increases as the shear rate increases. They are typically used in situations where higher borehole stability and higher drilling rates are required. Examples include bentonite, attapulgite, and hectorite.
    \end{enumerate}
\end{enumerate}
 
 Depending on the values of the yield point, $\tau_0$, and flow index, $n$, in Eqn. (\ref{HerBul}), drilling fluids may be classified as listed in the following table:
\begin{table}[h]
\begin{center}
\begin{TAB}(r,1cm,0.8cm)[2pt]{|c|c|c|c|}{|c|c|c|c|c|c|c|}
\textbf{S. no.} & \textbf{Fluid type} & \textbf{Yield point} & \textbf{Flow index}\\
1 & Shear thinning (pseudoplastic) & $\tau_0=0$ & $n<1$\\
2 & Newtonian & $\tau_0=0$ & $n=1$\\
3 & Shear thickening (dilatant) & $\tau_0=0$ & $n>1$\\
4 & Shear thinning with yield stress & $\tau_0>0$ & $n<1$\\
5 & Bingham plastic & $\tau_0>0$ & $n=1$\\
6 & Shear thickening with yield stress & $\tau_0>0$ & $n>1$\\
\end{TAB}
\end{center}
\caption{\label{fluid-type-1}Classification of drilling fluids based on their yield point and flow index.}
\end{table}

Up to this juncture, we have utilized rheological terminologies, specifically apparent viscosity, plastic viscosity, and yield point, on multiple occasions, without providing a comprehensive definition. We will address this deficiency in the subsequent section. Furthermore, we will discuss the methods for determining these parameters using Fann-35 and ZNN-D6 viscometers, which are presently widely used in the industry.

\section{Rheological properties of drilling fluids and their determination using rotational viscometers}
\label{sec: rheological-prop}
In drilling fluid rheology, apparent viscosity, plastic viscosity, and yield point are important parameters used to describe the flow behaviour of non-Newtonian (drilling) fluids. The mathematical definitions of these terms are as follows:
\begin{itemize}
    \item The apparent viscosity, $\mu_a$, is a measure of the resistance to flow of a non-Newtonian fluid under a specific set of conditions. It is defined as the ratio of (wall) shear stress to (wall) shear rate and is expressed in units of Pa$\cdot$s or cP. Mathematically, it can be written as
    \begin{equation}
        \mu_a = \frac{\tau_w}{\dot{\gamma}_w}.
        \label{appvis}
    \end{equation}

    \item The plastic viscosity, $\mu_p$, is a measure of the resistance to flow of a non-Newtonian fluid due to internal friction between its particles. It represents the slope of the shear stress-shear rate curve (also known as \textit{consistency plot}) in the laminar flow regime and is expressed in units of Pa$\cdot$s or cP. Mathematically, it can be written as:
    \begin{equation}
        \mu_p = \frac{\tau_{w,1} - \tau_{w,2}}{\dot{\gamma}_{w,1} - \dot{\gamma}_{w,2}},
        \label{plasvis}
    \end{equation}
    where $\dot{\gamma}_{w,1}$ and $\dot{\gamma}_{w,2}$ are arbitrary (wall) shear rates on the consistency plot (in the laminar flow regime), and $\tau_{w,1} = \tau_w(\dot{\gamma}_{w,1})$ and $\tau_{w,2} = \tau_w(\dot{\gamma}_{w,2})$. Combining Eqns. (\ref{HerBul}) and (\ref{plasvis}), the plastic viscosity can be reformulated as:
    \begin{equation}
        \mu_p = \eta\left(\frac{\dot{\gamma}_{w,1}^n - \dot{\gamma}_{w,1}^n}{\dot{\gamma}_{w,1} - \dot{\gamma}_{w,2}}\right).
        \label{plasvis2}
    \end{equation}

    \item The yield point, $\tau_0$, is the minimum amount of shear stress required to initiate flow in a non-Newtonian fluid. It represents the intercept of the shear stress-shear rate curve with the shear stress axis and is expressed in units of Pa or lb/100 sq. ft. Mathematically, as per Eqn. (\ref{HerBul}), the yield point is the shear stress at zero shear rate:
    \begin{equation}
        \tau_0 = \tau_w(\dot{\gamma}_w = 0).
        \label{yielpo}
    \end{equation}
\end{itemize}
In the petroleum industry, Fann-35 and ZNN-D6 viscometers are commonly used to measure the above-mentioned rheological properties of drilling fluids. 

The Fann-35 viscometer is a direct-indicating viscometer that operates based on the principle of rotational viscometry. It is a robust and reliable instrument that provides accurate and reproducible measurements of the rheological properties of drilling fluids. The instrument consists of a motor, a rotor, and a stator. The rotor is immersed in the drilling fluid, and the motor rotates the rotor at different speeds to induce shear in the fluid. The stator is fixed in place and acts as a stationary surface against which the fluid flows.

The Fann-35 viscometer is widely used due to its simple operation and well-established measurement protocols. It operates at two different rotational speeds, $300$ and $600$ rpm, and measures the torque required to rotate the spindle at these constant speeds through a fluid sample, which is proportional to the shear stress on the fluid. This allows for the calculation of the apparent viscosity, plastic viscosity, and yield point of the (drilling) fluid.

On the other hand, the ZNN-D6 viscometer is a coaxial cylinder rotational viscometer that can measure the rheological properties of drilling fluids at different speeds and shear rates. The instrument has a user-friendly interface and can be easily calibrated and maintained. It comprises a motor, a spindle, a torque sensor, a temperature sensor, and a digital display unit. The spindle is immersed in the fluid sample, and the motor rotates it at a constant speed. The torque sensor measures the resistance of the fluid to the spindle's rotation, and the temperature sensor records the temperature of the fluid during the measurement. 

The ZNN-D6 viscometer, similar to the Fann-35 viscometer, operates on the principle of measuring the torque required to rotate a spindle at a constant speed within a fluid sample. The shear stress and shear rate of the fluid are then calculated from the torque measurements, and from these measurements, the apparent viscosity, plastic viscosity, and yield point can be calculated. The procedure for measuring these rheological parameters is similar to that of the Fann-35 viscometer, as discussed earlier. However, the ZNN-D6 viscometer has a wider range of rotational speeds, can measure the rheological properties of fluids at high shear rates, has the ability to control temperature during measurement, and has higher accuracy. This allows for the calculation of more complex rheological properties such as the power-law exponent and consistency index of non-Newtonian fluids. The instrument can also be used to measure the thixotropic behaviour of drilling fluids, which is a crucial property in drilling operations. 

Overall, the choice between the Fann-35 and ZNN-D6 viscometers depends on the specific needs of the application and the desired level of accuracy and complexity in the rheological measurements. Their widespread use in the industry has led to the development of many mathematical models and correlations that can be used to predict the rheological behaviour of drilling fluids under different conditions. These models are essential for designing drilling operations and optimizing the performance of drilling fluids.

\cite{lam2014interpretation} discussed the importance of measuring the rheological properties of drilling fluids using viscometers such as the Fann-35 and ZNN-D6 viscometers. They also provided a detailed explanation of the mathematical models used to determine the rheological properties of drilling fluids using the Fann-35 viscometer. Specifically, the R1-B1-F1 combination of the Fann-35 model was discussed in order to determine the apparent viscosity, plastic viscosity, and yield point of drilling fluids. The method involves first testing the fluid at a speed of 300 rpm and recording the dial torque reading, $\theta_{300}$, and then at a speed of 600 rpm and recording the dial torque reading, $\theta_{600}$. The above-mentioned rheological properties are then calculated using the following set of equations:
\begin{dgroup}
\begin{dmath}
    \mu_a~(\text{cP}) = kf_s\frac{\theta_{N_1}}{N_1},
    \label{av}
\end{dmath}
\begin{dmath}
    \mu_p~(\text{cP}) = \frac{300}{N_3-N_2}(\theta_{N_3}-\theta_{N_2}),
    \label{pv}
\end{dmath}
\begin{dmath}
    \tau_0~(\text{Pa}) =
    \theta_{N_4} - \mu_p\frac{N_4}{300}.
    \label{yp}
\end{dmath}
\label{rheo-prop-set-1}
\end{dgroup}
In these equations, $N_i$ represents an arbitrarily chosen Fann rotational speed and $\theta_{N_i}$ is the corresponding dial torque reading ($i=1,2,3,4$). The constant $k$ represents the overall instrument constant, and $f_s$ is the torsion spring factor. For the standard rotor-bob combination R1-B1, $k$ is set to 300, and $f_s$ is set to 1 for the standard torsion spring F1. 

Setting $N_1=N_3=600$ rpm and $N_2=N_4=300$ rpm in Eqn. \eqref{rheo-prop-set-1}, and using the values of $k$ and $f_s$ for the standard rotor-bob combination R1-B1 and torsion spring F1, we obtain the following simplified set of equations:
\begin{dgroup}
\begin{dmath}
    \mu_a~(\text{cP}) = \frac{\theta_{600}}{2},
    \label{av_1}
\end{dmath}
\begin{dmath}
    \mu_p~(\text{cP}) = \theta_{600} - \theta_{300},
    \label{pv_1}
\end{dmath}
\begin{dmath}
    \tau_0~(\text{Pa}) = 2\theta_{300} - \theta_{600}.
    \label{yp_1}
\end{dmath}
\label{rheo-prop-set-2}
\end{dgroup}
These simplified equations are widely used in the oil and gas industry to determine the apparent viscosity, plastic viscosity, and yield point of drilling fluids using a Fann-35 viscometer.

In the subsequent chapter, a detailed discussion of our proposed mathematical framework will be presented. This framework is designed to offer more precise forecasts of the rheological characteristics of drilling fluids, surpassing the accuracy achieved by the pioneering research conducted by \cite{li2020rheological}.
\chapter{Mathematical framework}
\label{chap:2}
This chapter introduces a novel mathematical model developed for the precise determination of rheological properties in a diverse range of fluids, including both Newtonian and non-Newtonian fluids, with a particular focus on drilling fluids. The proposed model utilises the Marsh funnel as the primary equipment and is based on the well-established \textit{Herschel-Bulkley model}, complemented by mathematical models from previous research studies \cite{li2020rheological, sedaghat2017novel, sedaghat2016mathematical}. In line with the model proposed by \cite{li2020rheological}, as discussed in Chapter \ref{chap:1}, our developed model exhibits remarkable simplicity, enabling even individuals lacking expertise in the field to easily assess the rheological properties of fluids. The Marsh funnel serves as the sole equipment requirement, rendering the model independent of an external power source. Furthermore, its efficiency is highlighted by the absence of complex calculations and the minimal computational time required. Moreover, as discussed in Chapter \ref{chap:1}, our newly developed mathematical model offers a modest yet significant enhancement in the accuracy of determining various rheological properties of drilling fluids using the Marsh funnel, surpassing the accuracy achieved by the model proposed in \cite{li2020rheological}.

\begin{figure}[H]
    \centering
    \includegraphics[width=0.6\textwidth]{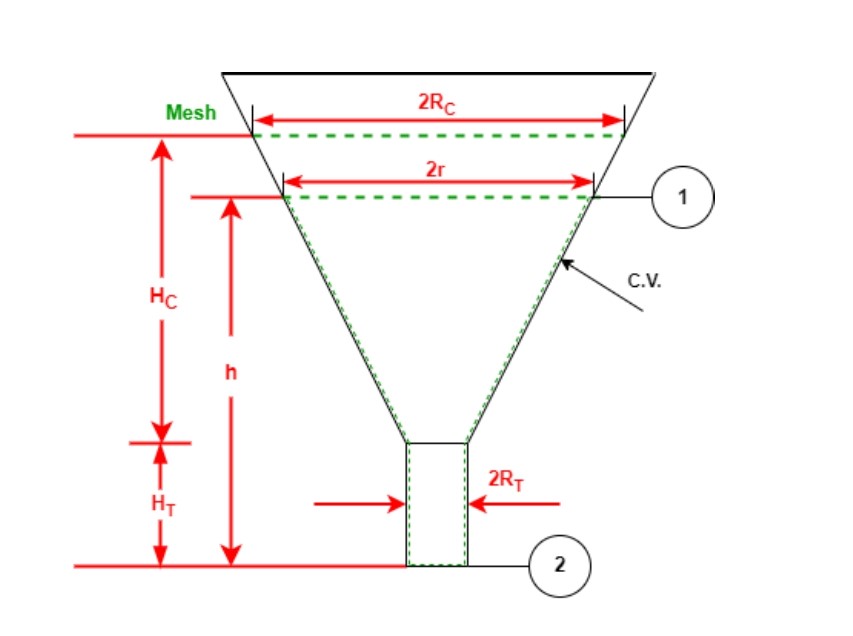}
    \caption{Schematic representation of the Marsh funnel's geometry.}
    \label{marsh}
\end{figure}

Before developing the mathematical model, it is essential to provide a detailed discussion on the geometry, properties, and terminologies associated with the Marsh funnel, which will be frequently referenced in this chapter and the subsequent one. The Marsh funnel serves as a cost-effective and widely employed device in the drilling industry for determining the rheological properties of drilling fluids. Fig. \ref{marsh} depicts the structure of the Marsh funnel, comprising two sections: an upper section in the form of an inverted conical frustum (truncated cone) and an attached lower section composed of a small cylindrical tube with a significantly smaller radius compared to the upper radius of the truncated cone. Typically, both sections are constructed using the same material, either plastic or metal, ensuring compatibility with various drilling fluids. To facilitate the separation of coarser particles from the drilling fluids, a circular sieve is affixed to the upper section of the Marsh funnel, as illustrated in Fig. \ref{marsh}, enabling the flow of finer fluid mixtures through the funnel. The dimensions of a standard Marsh funnel, as reported in \cite{sedaghat2017novel, li2020rheological}, are summarised in the following table:
\begin{table}[h]
\begin{center}
\begin{TAB}(r,1cm,0.6cm)[2pt]{|c|c|c|}{|c|c|c|c|c|}
\textbf{Dimension} & \textbf{Imperial (in)} & \textbf{Metric (cm)}\\
Radius of the truncated cone section at the sieve ($R_C$) & 2.75 & 6.985\\
Height of the truncated cone section at the sieve ($H_C$) & 11 & 27.94\\
Radius of the tube section ($R_T$) & $\frac{3}{32}$ & 0.238125\\
Height of the tube section ($H_T$) & 2 & 5.08
\end{TAB}
\end{center}
\caption{\label{marsh-dim}Dimensions of a standard Marsh funnel.}
\end{table}

After exploring the geometry of the Marsh funnel, we shift our focus to an essential terminology associated with this apparatus: the ``final discharge time,'' represented by $t_F$. This parameter signifies the duration required for the specimen fluid mixture, filled up to the sieve level of the Marsh funnel (reaching a height of $H_T + H_C$, with reference to the horizontal line ($2$) in Fig. \ref{marsh}), to fully empty the funnel. Captured using a stopwatch, this recorded discharge time assumes significant importance as one of the key variables for calculating the rheological properties of the fluid under examination using our proposed model, as elucidated in the forthcoming sections.

Furthermore, within the context of our study, the term \textit{fluid height}, denoted as $h$, designates the height to which the specimen fluid is filled within the Marsh funnel, with reference to the horizontal line ($2$) in Fig. \ref{marsh}. Consequently, the fluid height $h=0$ corresponds to the empty Marsh funnel or when the specimen fluid is completely drained out at time $t_F$. In a similar vein, the \textit{fluid radius}, represented by $r$, denotes the radius of the top free surface of the specimen fluid. Concurrently, the \textit{fluid volume}, denoted as $V_F$, signifies the total volume of the specimen fluid contained within the funnel.

With all the necessary prerequisites in place, we are prepared to embark on the development of our mathematical framework from its foundational principles.

\section{Formulating analytical relationships between fluid height, fluid radius, and fluid volume}
In this section, we will present a brief overview of the analytical approach employed by \cite{sedaghat2016mathematical} to formulate the relationship between fluid volume, fluid height, and fluid radius. Subsequently, we will demonstrate how the authors derived an expression for fluid volume solely as a function of fluid height, by expressing the fluid radius solely in terms of fluid height. This preliminary advancement serves as the initial building block of our mathematical model, forming a foundation for the subsequent development and analysis of the framework.

It is noteworthy to highlight that prior to the research conducted by \cite{sedaghat2016mathematical}, the Marsh funnel was commonly represented as an inverted complete cone rather than an inverted truncated cone coupled with a small cylindrical tube featuring a significantly smaller radius compared to the other radius of the truncated cone, as illustrated in Fig. \ref{marsh}. This modelling approach introduced a minor deviation in the determination of the fluid volume for the specimen drilling fluid within the Marsh funnel, which consequently emerged as a notable source of error in accurately assessing the rheological properties of the specimen fluid, as documented in studies such as \cite{balhoff2011rheological} and \cite{guria2013rheological}.

Furthermore, an additional substantial source of error stemmed from the neglect of fluid dynamics analysis within the tube section. \cite{sedaghat2017novel} demonstrated the significant contribution of analysing the fluid dynamics within this particular section, as it plays a pivotal role in precisely estimating the yield point and potential head. These parameters, in turn, are crucial for determining the rheological properties of the specimen drilling fluid.

Utilising the schematic representation of the Marsh funnel geometry presented in Fig. \ref{marsh}, \cite{sedaghat2016mathematical} derived the volume of the truncated conical section, denoted as $V_C$, and the volume of the tube section, denoted as $V_T$. The volume of the truncated conical section is given by
\begin{equation}
    V_C = \frac{1}{3}\pi(R_C^2 + R_CR_T + R_T^2)H_C,
    \label{consec}
\end{equation}
and the volume of the tube section is given by
\begin{equation}
    V_T = A_T H_T,
    \label{tubsec}
\end{equation} 
where $A_T = \pi R_T^2$ is the cross-sectional area of the tube section. Consequently, the total volume of the Marsh funnel, denoted as $V_M$, is determined by the sum of $V_C$ and $V_T$, as follows:
\begin{equation}
    V_M = V_C + V_T = \pi\left[\frac{1}{3}\left(R_C^2 + R_CR_T + R_T^2\right)H_C + R_T^2 H_T\right].
    \label{tot_vol}
\end{equation}

By substituting the dimensions of a standard Marsh funnel from Table \ref{marsh-dim} into Eqn. (\ref{tot_vol}), the total volume of the funnel is calculated. The computed value is approximately $1.479~\ell$ (with a more exact value of $1.47876864119958~\ell$). It is noteworthy that this calculated value exhibits a slight deviation from the reported value of $1.577~\ell$ in \cite{sedaghat2016mathematical}, where slightly different values of $R_C$ and $H_C$ were employed (i.e., $2.81\text{ in}$ and $11.25\text{ in}$, respectively) instead of the true values stated in Table \ref{marsh-dim} (i.e., $2.75\text{ in}$ and $11\text{ in}$, respectively).

To establish the relationship between fluid volume, fluid height, and fluid radius, consider a specimen fluid filled in the Marsh funnel up to a given fluid height, denoted as $h$, with the corresponding fluid radius expressed as a function of fluid height, denoted as $r(h)$. The fluid volume, represented by the control volume (C.V.) outlined with dashed lines in Fig. \ref{marsh}, can be defined as a function of fluid height and fluid radius as follows:
\begin{equation}
    V_F\left[h, r(h)\right] = \begin{cases}
    \pi r^2h & \text{ if }0\le h \le H_T,\\
    \pi\left[\frac{1}{3}\left(r^2 + R_Tr + R_T^2\right)(h-H_T) + R_T^2 H_T\right] & \text{ if } H_T\le h\le H_M.
    \end{cases}
    \label{cont.vol}
\end{equation} 
Here, $H_M = H_T + H_C$ represents the total height of the Marsh funnel up to the sieve, measured from the bottom end of the tube section.

By referring to the schematic representation of the Marsh funnel in Fig. \ref{marsh} and applying elementary geometric principles, we can derive the fluid surface radius as a function of fluid height. The relationship is given by:
\begin{equation}
    r(h) = \begin{cases}
    R_T & \text{ if }0\le h \le H_T,\\
    R_T + \left(\frac{R_C-R_T}{H_C}\right)(h-H_T) & \text{ if } H_T\le h\le H_M.
    \end{cases}
    \label{ra.he}
\end{equation}
To simplify the second expression in Eqn. \ref{ra.he}, we can rewrite it as a linear polynomial in $h$. Therefore, Eqn. \ref{ra.he} can be expressed as follows: 
\begin{equation}
    r(h) = \begin{cases}
    R_T & \text{ if }0\le h \le H_T,\\
    b_1h + b_2 & \text{ if } H_T\le h\le H_M.
    \end{cases}
    \label{ra.he2}
\end{equation}
Here, the coefficients $b_1$ and $b_2$ are defined as: $$b_1 = \frac{R_C-R_T}{H_C} \text{ and }b_2 = R_T - b_1H_T.$$

By substituting the expression for fluid radius as a linear polynomial in $h$ from Eqn. (\ref{ra.he2}) into Eqn. (\ref{cont.vol}), the fluid volume can be represented as a cubic polynomial in $h$. The expression is given by:
\begin{equation}
    V_F(h) = \begin{cases}
    A_Th & \text{ if }0\le h \le H_T,\\
    a_3h^3 + a_2h^2 + a_1h + a_0 & \text{ if } H_T\le h\le H_M.
    \end{cases}
    \label{cont.vol2}
\end{equation}
Here, the coefficients $a_3, a_2, a_1$, and $a_0$ are determined as follows: 
$$a_3 = \frac{1}{3}\pi b_1^2,~a_2 = \pi b_1b_2,~a_1 = \pi b_2^2,\text{ and }a_0 = \pi H_T\left[R_T^2 - \frac{1}{3}\left(b_2^2 + b_2R_T + R_T^2\right)\right].$$

The values of the coefficients of the cubic polynomial in Eqn. (\ref{cont.vol2}) can be determined using the dimensions of the standard Marsh funnel listed in Table \ref{marsh-dim}. The exact values of these coefficients are presented in Table \ref{coeffa}:
\begin{table}[H]
\begin{center}
\begin{TAB}(r,1cm,0.8cm)[2pt]{|c|c|}{|c|c|c|c|c|}
\textbf{Coefficient} & \textbf{Value}\\
$a_0$ (cm$^{3}$) & $-3.34334894814809$\\
$a_1$ (cm$^{2}$) & $3.07024556920872$\\
$a_2$ (cm) & $-0.749959403939138$\\
$a_3$ (dimensionless) & $0.0610634225480648$\\
\end{TAB}
\end{center}
\caption{\label{coeffa}Values of the cubic polynomial coefficients in Eqn. \eqref{cont.vol2}.}
\end{table}

To determine the total volume of the Marsh funnel, Eqn. \eqref{cont.vol2} is utilised by substituting $h=H_M$ and employing the coefficient values from Table \ref{coeffa}. Applying this approach, the total volume of the Marsh funnel is computed as:
\begin{equation*}
V_M = V_F(h=H_M) = a_3H_M^3 + a_2H_M^2 + a_1H_M + a_0 = 1478.76864119958 \text{ m}\ell
\approx 1.479~\ell.
\end{equation*}
This value aligns precisely with the total volume derived using Eqn. (\ref{tot_vol}), providing confirmation and validation of the mathematical framework proposed by \cite{sedaghat2016mathematical} for establishing correlations among fluid height, fluid radius, and fluid volume.

\section{Introducing the flow factor: Establishing its relationship with the final discharge time}
\label{sec: assumptions}
In this section, we establish a simple yet universal inverse linear relationship between the ``flow factor'', a characteristic parameter intrinsic to the specimen drilling fluid, and the final discharge time of the fluid. This relationship plays a pivotal role in determining the rheological properties of the fluid, making it a crucial element of our mathematical framework.

We initiate by presenting the mathematical framework proposed by \cite{sedaghat2016mathematical} to quantify the volumetric flow rate and the \textit{discharge time} of the specimen fluid within the Marsh funnel as a function of its fluid height. Moreover, the concept of the flow factor is introduced within this framework. Subsequently, we leverage this established framework to establish the aforementioned inverse linear relationship between the flow factor and the final discharge time.

The development of this mathematical framework is predicated upon the following assumptions:
\begin{enumerate}[label=(\alph*)]
    \item The flow of the drilling fluid specimen within the funnel is assumed to be one-dimensional, implying that the fluid speed, denoted by $v$, and the fluid pressure, denoted by $p$, solely vary in the direction of fluid flow, $z$, and with time, $t$. This assumption can be mathematically expressed as $v(x,y,z,t) = v(z,t)$ and $p(x,y,z,t) = p(z,t)$. This assumption holds true when the variations in fluid speed and fluid pressure along the $x$ and $y$ directions are considerably smaller compared to the variations along the $z$ direction, which is approximately the case for the flow of drilling fluid through a Marsh funnel. Furthermore, this assumption is justified by assumption (d), which postulates the presence of a flat free surface throughout the duration of the experiment.
    
    \item The fluid flow inside the Marsh funnel is incompressible. This assumption implies that the fluid density, $\rho$, remains constant in both space and time, and can be expressed as $\rho(x,y,z,t) = \rho$. This assumption is applicable when changes in fluid density caused by variations in pressure are negligible, as is the case in the present scenario.

   Moreover, considering an arbitrary control space defined by the Marsh funnel walls as side boundaries and the horizontal cross-sectional surfaces within the fluid as top and bottom boundaries, and considering an arbitrary infinitesimal time interval $[t, t + dt]$, we can deduce the following \textit{assuming no accumulation} (refer to assumption (c) for more details) within the control space:
    \begin{equation*}
        \dot{m}_{F,i}(t') - \dot{m}_{F,o}(t') = 0 \implies \dot{m}_{F,i}(t') = \dot{m}_{F,o}(t'),~t\le t' \le t+dt,
    \end{equation*}
    where $\dot{m}_{F,i}(t')$ and $\dot{m}_{F,o}(t')$ represent the input and output mass flow rates, respectively, at time $t'$ within the control space.  Consequently, it follows that the mass flow rate across all cross-sections within the fluid remains constant, denoted as $\dot{m}_{F}(z,t') = C \in \mathbb{R}_{\ge 0}$ for all $0\le z \le h$ and all $t \le t' \le t+ dt$. As a result, the volumetric flow rate, $Q_F(z,t') = C/\rho$, also remains constant across all cross-sections ($z\in [0,h]$) within the fluid during this time interval ($[t, t+dt]$).
    
    Hence, it can be concluded that the mass flow rate and the volumetric flow rate of the fluid are independent of the selected cross-section, although it may vary with time. In other words, $\dot{m}_F(z,t) = \dot{m}_F(t)$ and $Q_F(z,t) = Q_F(t)$ for all spatial coordinates $z\in [0,h]$ and all times $t > 0$.
    
    \item There is a continuous outflow of fluid from the Marsh funnel, and no fluid accumulation occurs within any arbitrarily selected control space (defined by two confining cross-sections of the fluid). When combining the latter part of this assumption with assumption (b), it can be inferred that the mass flow rate and volumetric flow rate of the fluid are independent of the chosen cross-section, albeit subject to variation with time, as discussed in the conclusion of assumption (b).
    
    Mathematically, the first part of this assumption implies that $\dot{m}_F(z=0,t)\neq 0$ for all time $t>0$. However, based on the conclusion of assumption (b), we have $0\neq \dot{m}_F(z=0,t) = \dot{m}_F(t) = \dot{m}_F(z,t)$ for all $0\le z \le h$ and all $t>0$. Consequently, it follows that $Q_F(t) = Q_F(z,t) \neq 0$ for all $0\le z \le h$ and all $t>0$, leading to the inference that the mass flow rate and the volumetric flow rate of the fluid in the Marsh funnel (across all cross-sections) are nonzero at any given time.
    
    Furthermore, due to the uniformity of the mass flow rate and the volumetric flow rate across all cross-sections at a given time, we may infer that
    \begin{subequations}\label{nonzero}
        \begin{align}
            -\dot{M}_F(t) &= \dot{m}_F(t) \neq 0, \text{ and} \label{mfrnonzero}\\
            -\dot{V}_F(t) &= Q_F(t) \neq 0, \label{vfrnonzero}
        \end{align}
    \end{subequations}
    for all time $t>0$, where $M_F(t)$ represents the total mass of the fluid present in the funnel (referred to as \textit{fluid mass}) at time $t$. This implies that both the fluid mass and fluid volume of the specimen fluid continuously decreases with time.
    
    \item The free top surface of the fluid in the Marsh funnel is flat consistently throughout the experiment. This particular assumption greatly simplifies the analysis of fluid flow dynamics inside the funnel. By considering the fluid surface as a single-level surface, a consistent and unique fluid height is established, eliminating the need to account for varying fluid heights at different points on the free surface. Consequently, this allows the fluid speed, $v$, and the fluid pressure, $p$, to be considered uniform across any given cross-section of the fluid within the funnel at any given time, as discussed in the conclusion of assumption (a): the variations in fluid speed and fluid pressure in the $x$ and $y$ directions are negligible compared to the variations in the $z$ direction. Collectively, these findings provide additional support for the assumption of one-dimensional flow as proposed in assumption (a).
\end{enumerate}

With the given assumptions in place, let us temporarily shift our focus from drilling fluids (viscous) and analyse the fluid flow of inviscid fluids in the Marsh funnel, implying fluids with negligible dynamic viscosity ($\mu \approx 0$). Consequently, for any given time $t \ge 0$, Bernoulli's principle for unsteady, inviscid, and incompressible fluid flow yields the following equation for all $z\in [0,h(t)]$, where $h(t)$ represents the fluid height at time $t$:
\begin{equation}
    \frac{1}{2}v^2\left(z,t\right) + gz + \frac{1}{\rho}p\left(z,t\right) = C(t),
    \label{bernoulli}
\end{equation}
where $C(t)\in \mathbb{R}$ is a time-dependent constant. Applying Eqn. \eqref{bernoulli} at the top free surface of the fluid (cross-section 1 in Fig. \ref{marsh}), we obtain:
\begin{equation}
    \frac{1}{2}v^2\left[z=h(t),t\right] + gh(t) + \frac{1}{\rho}p\left[z=h(t),t\right] = C(t).
    \label{bern1}
\end{equation}
Furthermore, by reapplying Eqn. \eqref{bernoulli} at the bottom surface of the tube section (cross-section 2 in Fig. \ref{marsh}), we arrive at:
\begin{equation}
    \frac{1}{2}v^2\left(z=0,t\right) + g(0) + \frac{1}{\rho}p\left(z=0,t\right) = C(t).
    \label{bern2}
\end{equation}

Note that the velocity at the top free surface of the fluid is nearly negligible, allowing us to set $v[h(t), t] = 0$. Moreover, throughout the experiment, the volumetric flow rate of the fluid remains relatively low. Consequently, we can assume that the pressure acting on the fluid at both the top free surface and the bottom surface of the tube section, at any given time, is in equilibrium with the surrounding atmospheric pressure. This assumption leads to $p[h(t), t] = p_{atm} = p(0,t)$. Additionally, $v(0,t)$ represents the \textit{exit velocity} at time $t$, denoted as $v_e(t)$. By substituting these values into Eqns. \eqref{bern1} and \eqref{bern2}, and subsequently comparing them, we arrive at the following equation:
\begin{equation}
    gh(t) + \frac{1}{\rho}p_{atm} = C(t) = \frac{1}{2}v_e^2(t) + \frac{1}{\rho}p_{atm}.
    \label{ber1}
\end{equation}
Upon simplification and rearrangement, we obtain:
\begin{equation}
    v_e(t) = \sqrt{2gh(t)}.
    \label{exvel}
\end{equation}
By removing the time dependency from both sides of Eqn. \eqref{exvel}, we arrive at:
\begin{equation}
    v_e(h) = \sqrt{2gh}.
    \label{exvel1}
\end{equation}
This reveals that the exit velocity of unsteady, inviscid, and incompressible fluid flows from a Marsh funnel is solely dependent on the fluid height.
    
In the context of unsteady and incompressible flows of viscous fluids (e.g., drilling fluids) in a Marsh funnel, it has been observed that the exit velocity is solely governed by the fluid height, similar to what has been established for inviscid fluids. However, it should be noted that the exit velocity, in this case, deviates from the exact equation stated in Eqn. \eqref{exvel}. To accurately quantify the exit velocity of viscous fluid flows in a Marsh funnel, particularly drilling fluid flows, \cite{sedaghat2016mathematical} proposed a modification to the expression for exit velocity in Eqn. \eqref{exvel1}. This modification involves multiplying the right-hand side of the equation by a dimensionless flow coefficient, known as the flow factor, represented by $f$. As a result, the modified equation for exit velocity is presented as follows:
    \begin{equation}
        v_e(h) = f\sqrt{2gh}.
        \label{exvelmod}
    \end{equation}

In the context of Newtonian viscous fluids, the flow factor $f$ remains constant over time. However, in the case of non-Newtonian viscous fluids, the flow factor becomes a time-dependent quantity due to its dependence on various time-varying fluid properties. Consequently, the flow factor is expressed as a function of these fluid properties. To determine specific values of $f$ for different combinations of these fluid properties, experimental measurements are performed at carefully selected combinations of these parameters. Subsequently, suitable curve fitting techniques are applied to establish a functional relationship. Nevertheless, in our study, while acknowledging the time-varying nature of most of the inherent properties of a given non-Newtonian drilling fluid, we assume the flow factor to remain approximately constant over time to facilitate ease of analysis.

Recall that the volumetric flow rate in the Marsh funnel at any given time $t\ge 0$ is equal to the volumetric flow rate at the exit of the funnel at that specific time. This relationship can be expressed as follows:
\begin{equation}
    Q_F(t) = Q_F(z=0, t) = A_T v_e(t).
    \label{volex}
\end{equation}
By substituting the expression for $Q_F(t)$ from Eqn. \eqref{vfrnonzero} and the expression for $v_e(t)$ from Eqn. \eqref{exvel} into Eqn. \eqref{volex}, we arrive at:
    \begin{equation}
        -\dot{V}_F(t) = A_T f\sqrt{2gh(t)},
    \end{equation}
for all times $t\ge 0$. Applying the chain rule of differentiation to the left-hand side of Eqn. \eqref{volex} and then multiplying both sides by $-1$, results in:
    \begin{equation}
        \frac{dV_F}{dh}\frac{dh}{dt} = -A_T f\sqrt{2gh}.
    \end{equation}
Upon multiplying both sides by $dh/dV_F$, we obtain:
\begin{equation}
    \frac{dh}{dt} = -A_T f\sqrt{2gh}\frac{dh}{dV_F}.
    \label{dh/dt}
\end{equation}
This equation represents the relationship between the change in fluid height $h$ with respect to time $t$ and the change in fluid volume $V_F$ with respect to fluid height.

By differentiating both sides of Eqn. \eqref{cont.vol} with respect to the fluid height $h$, we arrive at the following expression:
\begin{equation}
    \frac{dV_F}{dh} = 
    \begin{cases}
    A_T & \text{if } 0\le h \le H_T,\\
    3a_3h^2 + 2a_2h + a_1 & \text{if } H_T\le h \le H_M.
    \end{cases}
    \label{dv/dh}
\end{equation}
Next, we calculate the inverse of both sides of Eqn. \eqref{dv/dh} to derive the expression for $dh/dV_F$. Substituting this expression in Eqn. \eqref{dh/dt}, we obtain:
\begin{equation}
    \frac{dh}{dt} = 
    \begin{cases}
    -f\sqrt{2gh} & \text{if } 0\le h \le H_T,\\
    -A_T f \sqrt{2g}\frac{\sqrt{h}}{3a_3h^2 + 2a_2h + a_1} & \text{if } H_T\le h \le H_M.
    \end{cases}
    \label{dh/dt2}
\end{equation}
We then invert both sides of Eqn. \eqref{dh/dt2} and multiply both sides of the resulting equation by $dh$ to obtain the following expression for the time differential $dt$:
\begin{equation}
    dt = 
    \begin{cases}
    -\frac{1}{f\sqrt{2g}}h^{-0.5} dh & \text{if } 0\le h \le H_T,\\
    -\frac{1}{A_T f \sqrt{2g}}\left(3a_3 h^{1.5} + 2a_2h^{0.5} + a_1h^{-0.5}\right) dh & \text{if } H_T\le h \le H_M.
    \end{cases}
    \label{dt}
\end{equation}

Suppose that, initially, the specimen fluid in the Marsh funnel is filled up to the sieve of the funnel, leading to the initial condition $t(h=H_M) = 0$, and let the time taken to reach an arbitrary fluid height $h=H \in [0, H_M]$ be $t=\tau(H)$. By integrating the left-hand side from $t=0$ to $t=\tau(H)$ and the right-hand side from $h=H_M$ to $h = H$ of Eqn. \eqref{dt}, we obtain the expression for the time required to change the fluid height from $h = H_M$ to $h = H$ in terms of $H$:
\begin{dmath}
\begin{aligned}
    \tau(H) &= \int_0^{\tau(H)} dt = 
    \begin{cases}
    -\beta \left[\displaystyle\int_{H_M}^{H_T}\left(\sum_{j=1}^3 ja_jh^{j-1.5}\right) dh + A_T\displaystyle\int_{H_T}^{H} \left(h^{-0.5}\right)dh\right] & \text{if } 0\le H \le H_T,\\
    -\beta \displaystyle\int_{H_M}^{H}\left(\sum_{j=1}^3 ja_jh^{j-1.5}\right) dh & \text{if } H_T\le H \le H_M,
    \end{cases} \\
    &=
    \begin{cases}
    \beta \left[A_T\displaystyle\int_{H}^{H_T} \left(h^{-0.5}\right)dh + \sum_{j=1}^3\left(ja_j\displaystyle\int_{H_T}^{H_M} h^{j-1.5} dh\right)\right] & \text{if } 0\le H \le H_T,\\
    \beta \displaystyle\sum_{j=1}^3\left(ja_j\displaystyle \int_{H}^{H_M} h^{j-1.5} dh\right) & \text{if } H_T\le H \le H_M,
    \end{cases} \\
    &= 
    \begin{cases}
    -2\beta A_TH^{0.5} + \beta\left[2A_TH_T^{0.5} + \displaystyle\sum_{j=1}^3\left(\frac{ja_j}{j-0.5}\left[H_M^{j-0.5} - H_T^{j-0.5}\right] \right)\right] & \text{if } 0\le H \le H_T,\\
    -\beta\displaystyle\sum_{j=1}^3\left(\frac{ja_j}{j-0.5}H^{j-0.5}\right) + \beta\displaystyle\sum_{j=1}^3\left(\frac{ja_j}{j-0.5}H_M^{j-0.5}\right) & \text{if } H_T\le H \le H_M,
    \end{cases}
    \label{t}
\end{aligned}
\end{dmath}
where $\beta = \frac{1}{A_T f \sqrt{2g}}$ is a real constant. We can define the following real constants:
\begin{align*}
\begin{gathered}
    \alpha_j = -\beta\frac{ja_j}{j-0.5},~j=1,2,3,~\beta_1 = -2\beta A_T,\\
    \beta_2 = -\displaystyle\sum_{j=1}^3\left(\alpha_j H_M^{j-0.5}\right), \text{ and } \beta_3 = -\beta_1 H_T^{0.5} + \beta_2 + \displaystyle\sum_{j=1}^3\left(\alpha_jH_T^{j-0.5}\right).
\end{gathered}
\end{align*}
With these constants, Eqn. \eqref{t} can be simplified as follows:
\begin{equation}
    \tau(H) =
    \begin{cases}
        \beta_1 H^{0.5} + \beta_3 & \text{if } 0\le H \le H_T,\\
        \displaystyle\sum_{j=1}^3 \left(\alpha_j H^{j-0.5}\right) + \beta_2 & \text{if } H_T\le H \le H_M.
    \end{cases}
    \label{t1}
\end{equation}
For brevity, we shall refer to $\tau(H)$ as the \textit{fluid height transition time}, or simply the \textit{transition time}.

An important term closely related to the transition time is the ``discharge time.'' Given a specimen fluid in the Marsh funnel with a fluid height $H\in [0, H_M]$, the discharge time, denoted by $t_D(H)$, is defined as the time required to completely drain the fluid present in the funnel, i.e., change the fluid height from $h = H$ to $h = 0$. The discharge time can be expressed as follows:
\begin{dmath}
\begin{aligned}
    t_D(H) &= \tau(0) - \tau(H) =
    \begin{cases}
        -\beta_1 H^{0.5} & \text{if } 0\le H \le H_T,\\
        \beta_3-\beta_2-\displaystyle\sum_{j=1}^3\left(\alpha_j H^{j-0.5}\right) & \text{if } H_T\le H \le H_M,
    \end{cases}
    \label{td}
\end{aligned}
\end{dmath}
Here, $\tau(0)$ and $\tau(H)$ represent the transition times corresponding to fluid heights $h = 0$ and $h = H$, respectively.

Having established the concepts of transition time and discharge time, and developed their expressions as functions of the fluid height, we can now delve into the expression for the final discharge time, denoted as $t_F$. We can obtain $t_F$ by substituting $H = 0$ in Eqn. \eqref{t1} or $H = H_M$ in Eqn. \eqref{td}, leading to:
\begin{equation}
        t_F = 
        \begin{cases}
            \tau(0) = \beta_3,\\
            t_D(H_M) = \beta_3-\beta_2-\displaystyle\sum_{j=1}^3\left(\alpha_j H_M^{j-0.5}\right) = \beta_3 -\beta_2+\beta_2 = \beta_3.
        \end{cases}
\label{tF}
\end{equation}
Since $\beta_3$ is a real constant determined solely by the flow factor $f$ and the dimensions of the Marsh funnel, $t_F$ also becomes a real constant reliant solely on $f$ and the funnel's dimensions. We now establish the relationship between the flow factor and the final discharge time:
\begin{align*}
    t_F = \beta_3 &= -\beta_1 H_T^{0.5} + \beta_2 + \displaystyle\sum_{j=1}^3\left(\alpha_jH_T^{j-0.5}\right)\\
    &= \beta\left[2A_TH_T^{0.5} + \displaystyle\sum_{j=1}^3\left(\frac{ja_j}{j-0.5}\left[H_M^{j-0.5} - H_T^{j-0.5}\right] \right)\right]\\
    &= \frac{1}{A_Tf\sqrt{2g}}\left[2A_TH_T^{0.5} + \displaystyle\sum_{j=1}^3\left(\frac{ja_j}{j-0.5}\left[H_M^{j-0.5} - H_T^{j-0.5}\right] \right)\right].
\end{align*}
By defining the constant $\alpha$ as:
\begin{equation}
    \alpha = \frac{1}{A_T\sqrt{2g}}\left[2A_TH_T^{0.5} + \displaystyle\sum_{j=1}^3\left(\frac{ja_j}{j-0.5}\left[H_M^{j-0.5} - H_T^{j-0.5}\right] \right)\right],
    \label{alpha}
\end{equation}
we can express the relationship between the flow factor and the final discharge time as:
\begin{equation}
    \boxed{f = \frac{\alpha}{t_F}}.
    \label{tF1}
\end{equation}
This relationship demonstrates an inverse linear dependency of the flow factor $f$ on the final discharge time $t_F$, where the constant $\alpha$ serves as the proportionality factor. A notable observation is that $\alpha$ is solely contingent on the dimensions of the Marsh funnel and not on any specimen fluid properties. As a result, it can be considered an inherent constant of the funnel, and henceforth, we shall refer to it as the \textit{Marsh funnel constant}. 

Furthermore, for any given specimen fluid, the final discharge time $t_F$ can be easily measured through experimentation. Once the final discharge time is determined, we can substitute its value into Eqn. \eqref{tF1} to readily obtain the flow factor of the specimen fluid. This ease of determining the flow factor (an intrinsic property of the specimen fluid that allows us to assess the rheological properties of the fluid effortlessly, as we will explore in the forthcoming sections of this chapter) underscores the significance and practical value of this relationship in our research.

In the forthcoming sections, we shall delve into the derivation of the expressions of crucial drilling fluid properties that are pertinent to the study of fluid dynamics within a confined space, specifically the Marsh funnel. These properties include head loss, average wall shear stress, and average wall shear rate. These novel expressions were originally formulated by \cite{sedaghat2017novel} and \cite{li2020rheological}. Subsequently, we will present the derivation of rheological properties for the three types of drilling fluids discussed in Chapter \ref{chap:1}: (i) Newtonian, (ii) non-weighted non-Newtonian, and (iii) weighted non-Newtonian. These rheological properties will be expressed as functions of fluid density $\rho$ and flow factor $f$, as developed by \cite{li2020rheological}.

As the culminating part of our mathematical framework, we will incorporate the expression for the flow factor, as given in Eqn. \eqref{tF1}, into these equations. This will lead us to the final expressions of the rheological properties in terms of fluid density and final discharge time $t_F$. These newly derived equations will play a pivotal role in Chapter \ref{chap:3}, where we will determine the rheological properties of the selected drilling fluids through the straightforward measurement of their final discharge times $t_F$ and the subsequent calculation of their densities.

\section{The head loss and the average wall shear stress}
\label{sec: head loss}
The precise determination of the average wall shear stress, $\langle\tau_w\rangle$, and average wall shear rate, $\langle\dot{\gamma}_w\rangle$, for a given specimen drilling fluid inside the Marsh funnel at fluid height $h$ is of utmost importance in determining the fluid's rheological properties. In this context, let $\tau_w(x,y,z,h)$ and $\dot{\gamma}_w(x,y,z,h)$ represent the wall shear stress and wall shear rate, respectively, at the point $(x,y,z)\in S(h)$, where $S(h)$ denotes the surface of the Marsh funnel between $z=0$ and $z=h$. Given that we have assumed one-dimensional flow, it follows that $\tau_w(x,y,z,h) = \tau(z,h)$ and $\dot{\gamma}_w(x,y,z,h) = \dot{\gamma}_w(z,h)$ for all $x$, $y$, $z$, and $h$. Moreover, let $A_{S(h)}$ denote the surface area of $S(h)$. We can then determine the average wall shear stress and the average wall shear rate as follows:
\begin{subequations}
\begin{align}
    \langle\tau_w\rangle(h) = \frac{1}{A_{S(h)}}\iint_{S(h)}\tau_w(x,y,z,h)~dS = \frac{1}{A_{S(h)}}\iint_{S(h)}\tau_w(z,h)~dS \label{tauw1},\\
    \langle\dot{\gamma}_w\rangle(h) = \frac{1}{A_{S(h)}}\iint_{S(h)}\dot{\gamma}_w(x,y,z,h)~dS = \frac{1}{A_{S(h)}}\iint_{S(h)}\dot{\gamma}_w(z,h)~dS \label{gammaw1}.
\end{align}
\label{shear}
\end{subequations}

The integrals in Eqn. \eqref{shear} can be evaluated using appropriate analytical or numerical techniques. However, obtaining explicit expressions for $\tau_w(z,h)$ and $\dot{\gamma}_w(z,h)$ for all $0\leq z \leq h$ and $0\leq h \leq H_M$ is a challenging task and remains an open area of research in computational fluid dynamics (CFD). Consequently, instead of adopting this direct approach to compute $\langle\tau_w\rangle(h)$ and $\langle\dot{\gamma}_w\rangle(h)$, we present the alternative method employed by \cite{sedaghat2017novel} for their determination. This alternative method does not directly compute $\langle\tau_w\rangle(h)$ and $\langle\dot{\gamma}_w\rangle(h)$; rather, it first focuses on establishing the expression for head loss and subsequently employs it to deduce the average wall shear stress and the average wall shear rate.

The second form of Bernoulli's equation, governing unsteady, inviscid, and incompressible fluid flows, is derived by dividing the first form, as given in Eqn. \eqref{bernoulli}, by the acceleration due to gravity $g$. This manipulation yields a concise expression that combines the three \textit{energy heads} in fluid dynamics: (i) the \textit{velocity head} $\frac{v^2(z,t)}{2g}$, (ii) the \textit{elevation head} $z$, and (iii) the \textit{pressure head} $\frac{p(z,t)}{\rho g}$. The resulting equation is presented as follows:
\begin{equation}
    \frac{v^2(z,t)}{2g} + z + \frac{p(z,t)}{\rho g} = C_1(t),
    \label{bernoulli2}
\end{equation}
where $C_1(t) = \frac{C(t)}{g}\in \mathbb{R}$ represents a time-dependent constant referred to as the \textit{total head}. The total head serves as a comprehensive measure of the fluid's mechanical energy per unit weight, incorporating contributions from its pressure, velocity, and elevation above a specified reference plane (often denoted as $z = 0$).

In our system, Eqn. \eqref{bernoulli2} cannot be directly applied because drilling fluids, as discussed in the previous section, exhibit substantial viscosity. As they flow through the Marsh funnel, they experience energy losses due to various factors, such as friction between the fluid and funnel walls, inter-layer friction, turbulence primarily at the entrance and exit of the tube section, and other sources leading to energy dissipation. All these energy losses are collectively known as ``head loss,'' which must be taken into account when modifying Bernoulli's equation (Eqn. \eqref{bernoulli2}) to apply it to real fluid flow systems, such as in our case.

Thus, the modified Bernoulli's equation applicable at an arbitrary time $t \ge 0$ and arbitrary cross-sections of the flowing drilling fluid in the Marsh funnel, defined by $z = z_1$ and $z = z_2$ where $0 \le z_1, z_2 \le h(t)$, can be expressed as follows:
\begin{equation}
    \frac{v^2(z_1,t)}{2g} + z_1 + \frac{p(z_1,t)}{\rho g} + h_L(z_1, z_2, t) = \frac{v^2(z_2,t)}{2g} + z_2 + \frac{p(z_2,t)}{\rho g},
    \label{mod-bernoulli}
\end{equation}
where $h_L(z_1, z_2, t)$ denotes the head loss between the specified cross-sectional planes. Specifically, the head loss can be calculated using the expression:
\begin{equation}
    h_L(z_1, z_2, t) = \frac{v^2(z_2,t) - v^2(z_1,t)}{2g} + (z_2-z_1) + \frac{p(z_2,t) - p(z_1,t)}{\rho g}.
    \label{head-loss}
\end{equation}

By setting $z_1 = h(t)$ and $z_2 = 0$, and recalling the expressions discussed in the previous section for the velocities and pressures at these locations, namely $v(z_1,t) = 0$, $v(z_2,t) = v_e(t)$, and $p(z_1,t) = p_{atm} = p(z_2,t)$, we can now evaluate the \textit{total head loss} across the two ends of the flowing fluid in the funnel at any time $t \ge 0$. This head loss, denoted as $h_T(t)$, is thus given by the following equation:
\begin{equation}
    h_T(t) = -h_L[h(t), 0, t)] = -\left[\frac{v_e^2(t)}{2g} - h(t)\right] = h(t)[1-f^2],
    \label{tot-head-loss-1}
\end{equation}
where $h(t=0) = H_M$. 

Alternatively, the total head loss can be determined in terms of the \textit{Darcy's friction factor}, also known as the \textit{Fanning friction factor} or simply the \textit{friction factor}, denoted by $f_D$. This dimensionless quantity, at any given time $t\geq 0$, represents the ratio of the total head loss, $h_T(t)$, to the exit velocity head, $\frac{v_e^2(t)}{2g}$, in a confined flow system, such as fluid flow through a Marsh funnel or pipe flow. Moreover, it serves as a measure of the average wall shear stress, $\langle\tau_w\rangle [h(t)]$. Specifically, for the case of fluid flow through the Marsh funnel, it is expressed as a function of time as follows \cite{sedaghat2017novel}:
\begin{equation}
    f_D(t) = \frac{8\langle\tau_w\rangle [h(t)]}{\rho v_e^2(t)},
    \label{friction-factor}
\end{equation}
for all time $t\ge 0$. In terms of the friction factor, \cite{sedaghat2017novel} derived the expression for the total head loss at any time $t\geq 0$ as follows:
\begin{equation}
    h_T(t) = f_D(t)\frac{H_T}{2R_T}\frac{v_e^2(t)}{2g} = \frac{H_T}{4gR_T}f_D(t)v_e^2(t).
    \label{tot-head-loss-2}
\end{equation}
By substituting the expression of $f_D(t)$ from Eqn. \eqref{friction-factor} into Eqn. \eqref{tot-head-loss-2}, the total head loss at any time $t\geq 0$ can be expressed in terms of the average wall shear rate at that time, $\langle \tau_w\rangle[h(t)]$, as follows:
\begin{equation}
    h_T(t) = \frac{2H_T}{\rho g R_T}\langle \tau_w\rangle[h(t)],
    \label{tot-head-loss-3}
\end{equation}
signifying that the total head loss, in terms of the fluid height $h\in [0, H_M]$, can be expressed as follows:
\begin{equation}
    h_T(h) = \frac{2H_T}{\rho g R_T}\langle \tau_w\rangle(h).
    \label{tot-head-loss-4}
\end{equation}

Upon comparing the expressions of $h_T(t)$ in Eqns. \eqref{tot-head-loss-1} and \eqref{tot-head-loss-3}, the following equation emerges:
\begin{equation}
    h(t)[1-f^2] = h_T(t) = \frac{2H_T}{\rho g R_T}\langle \tau_w\rangle[h(t)],
\end{equation}
from which we can deduce an algebraic function for $\langle \tau_w \rangle$ in terms of $h$ as follows:
\begin{equation}
    \langle \tau_w \rangle(h) = \frac{1}{2}\rho g h(1-f^2)\frac{R_T}{H_T}.
    \label{shear-stress-2}
\end{equation}
This equation represents the precise expression for the average wall shear stress, as derived by \cite{sedaghat2017novel}. It is noteworthy that Eqn. \eqref{shear-stress-2} demonstrates the average wall shear stress to be a linear function of the fluid height for any class of drilling fluid.

In the forthcoming section, we employ Eqn. \eqref{shear-stress-2} to establish the expression, originally derived by \cite{sedaghat2017novel} and \cite{li2020rheological}, for the average wall shear rate, $\langle \dot{\gamma}_w\rangle$. This expression is formulated as an algebraic function of the fluid height, $h$, and is specifically developed separately for Newtonian, non-weighted non-Newtonian, and weighted non-Newtonian drilling fluids.

\section{The average wall shear rate}
Before proceeding to derive the expression of the average wall shear rate, $\langle \dot{\gamma}_w\rangle$, in terms of the fluid height, $h$, for the three types of drilling fluids, it is essential to introduce the concept of the \textit{flow behaviour index}, denoted by $n'$. It should be emphasized that this index is distinct from the flow index, $n$, in Eqn. \eqref{HerBul}. For any given specimen drilling fluid, \cite{brydson1970flow} observed that the plot between the variables $\log\left(4Q_F/\pi R_T^3\right)$ and $\log \langle \tau_w \rangle$ forms a straight line. They defined the flow behaviour index as the inverse of the slope of this line. Consequently, the flow behaviour index can be expressed as follows:
\begin{equation}
    n' = \frac{d \log \langle\tau_w\rangle}{d \log \left(\frac{4Q_F}{\pi R_T^3}\right)}.
    \label{flow-index-1}
\end{equation}
Now, considering
$$d \log \langle\tau_w\rangle = \frac{1}{\ln 10}\frac{d\langle\tau_w\rangle}{\langle\tau_w\rangle}, \text{ and }d \log \left(\frac{4Q_F}{\pi R_T^3}\right) = \frac{1}{\ln 10}\frac{dQ_F}{Q_F},$$
we have
\begin{dmath}
\begin{aligned}
    n' = \frac{d \log \langle\tau_w\rangle}{d \log \left(\frac{4Q_F}{\pi R_T^3}\right)} &= \frac{Q_F}{\langle\tau_w\rangle}\frac{d\langle\tau_w\rangle}{dQ_F}\\
    &= \frac{A_T f\sqrt{2g}}{\frac{1}{2}\rho g(1-f^2)\frac{R_T}{H_T}}\frac{\sqrt{h}}{h}\frac{\frac{1}{2}\rho g(1-f^2)\frac{R_T}{H_T}}{A_Tf\sqrt{2g}}\frac{dh}{d\sqrt{h}}\\
    &= \frac{1}{\sqrt{h}}\frac{dh}{d\sqrt{h}} = \frac{1}{\sqrt{h}}{2\sqrt{h}} = 2.
    \label{flow-index-2}
\end{aligned}
\end{dmath}

In the context of Newtonian (drilling) fluids, \cite{sedaghat2017novel} proposed the following expression for the average wall shear rate in terms of the fluid height:
\begin{equation}
    -\langle\dot{\gamma}_w\rangle(h) = \frac{1}{2}\left(\frac{h}{H_T}\right)^{\frac{1}{2}}\frac{3n'+ 1}{4n'} \frac{4Q_F}{\pi R_T^3},
    \label{newton-shear-rate-1}
\end{equation}
where $Q_F$ represents the volumetric flow rate in the funnel at time $t = \tau(h)$. Now, considering $A_T = \pi R_T^2$, and substituting $n'=2$ and $Q_F = A_Tf\sqrt{2gh}$ into Eqn. \eqref{newton-shear-rate-1}, the above expression can be simplified as follows:
\begin{dmath}
\begin{aligned}
    -\langle\dot{\gamma}_w\rangle(h) &= \frac{1}{2A_TR_T\sqrt{H_T}}\left(3+\frac{1}{n'}\right)Q_Fh^{\frac{1}{2}}\\
    &= \frac{7A_Tf\sqrt{2g}}{4A_TR_T\sqrt{H_T}}h = \frac{7\sqrt{2g}}{4R_T\sqrt{H_T}}fh.
\end{aligned}
\label{newton-shear-rate-2}
\end{dmath}
Eqn. \eqref{newton-shear-rate-2} highlights that for Newtonian fluids, similar to the average wall shear stress, the average wall shear rate also follows a linear relationship with the fluid height.

For the category of non-weighted non-Newtonian drilling fluids, \cite{sedaghat2017novel} proposed the following expression for the average wall shear rate in terms of the fluid height:
\begin{equation}
    -\langle\dot{\gamma}_w\rangle(h) = \frac{1}{2}\left(\frac{h}{H_T}\right)\frac{3n'+ 1}{4n'} \frac{4Q}{\pi R_T^3}.
    \label{non-weight-non-newton-shear-rate-1}
\end{equation}
By employing the same approach as before, this equation can be simplified as follows:
\begin{dmath}
    \begin{aligned}
        -\langle\dot{\gamma}_w\rangle(h) 
        &= \frac{1}{2A_TR_TH_T}\left(3+\frac{1}{n'}\right)Q_Fh\\
        &= \frac{7A_Tf\sqrt{2g}}{4A_TR_TH_T}h^{\frac{3}{2}} = \frac{7\sqrt{2g}}{4R_TH_T}fh^{\frac{3}{2}}.
    \end{aligned}
    \label{non-weight-non-newton-shear-rate-2}
\end{dmath}

Lastly, moving on to the class of weighted non-Newtonian drilling fluids, \cite{li2020rheological} proposed the following expression for the average wall shear rate with respect to the fluid height:
\begin{equation}
    -\langle\dot{\gamma}_w\rangle(h) = \frac{1}{4\sqrt{2}}\left(\frac{h}{H_T}\right)\frac{3n'+ 1}{4n'} \frac{4Q}{\pi R_T^3},
    \label{weight-non-newton-shear-rate-1}
\end{equation}
This expression can be simplified in a manner similar to the previous cases, yielding:
\begin{dmath}
    \begin{aligned}
        -\langle\dot{\gamma}_w\rangle(h) 
        &= \frac{1}{4\sqrt{2}A_TR_TH_T}\left(3+\frac{1}{n'}\right)Q_Fh\\
        &= \frac{7A_Tf\sqrt{2g}}{8\sqrt{2}A_TR_TH_T}h^{\frac{3}{2}} = \frac{7\sqrt{g}}{8R_TH_T}fh^{\frac{3}{2}}.
    \end{aligned}
    \label{weight-non-newton-shear-rate-2}
\end{dmath}

An interesting observation from Eqns. \eqref{non-weight-non-newton-shear-rate-2} and \eqref{weight-non-newton-shear-rate-2} is that the average wall shear rate for both non-weighted and weighted non-Newtonian drilling fluids is proportional to $h^{\frac{3}{2}}$, hence indicating a non-linear relationship with $h$. This behaviour contrasts with the case of Newtonian drilling fluids, where the average wall shear rate exhibited a linear relationship with the fluid height.

\section{Rheological properties}
In Section \ref{sec: classification}, it was established that Newtonian (drilling) fluids exhibit a linear relationship between the (wall) shear stress, $\tau_w$, and the (wall) shear rate, $\dot{\gamma}_w$, described by Eqn. \eqref{Newtondrill}. Specifically, for a given specimen Newtonian drilling fluid in a Marsh funnel with fluid height $h$, this linear relationship can be expressed as follows:
\begin{equation}
    \tau_w(z,h) = \tau_w(x,y,z,h) = \mu \dot{\gamma}(x,y,z,h) = \mu \dot{\gamma}(z,h),
    \label{Newtondrill-2}
\end{equation}
for all $h\in [0,H_M]$ and any point $(x,y,z)\in S(h)$ (refer to Section \ref{sec: head loss} for the definition of $S(h)$). Here, $\mu$ denotes the dynamic viscosity of the Newtonian fluid under examination, which remains constant at all points along $S(h)$ for any $h\in [0,H_M]$.

By integrating Eqn. \eqref{Newtondrill-2} over the surface $S(h)$ and dividing by its surface area $A_{S(h)}$, we arrive at the following equation:
$$\frac{1}{A_{S(h)}}\iint_{S(h)} \tau_w(z,h)~dS = \frac{1}{A_{S(h)}}\iint_{S(h)} \mu\dot{\gamma}_w(z,h)~dS = \frac{\mu}{A_{S(h)}}\iint_{S(h)}\dot{\gamma}_w(z,h)~dS.$$
This equation establish a linear relationship between the average wall shear stress and the average wall shear rate:
\begin{equation}
    \langle \tau_w \rangle(h) = \mu\langle \dot{\gamma}_w \rangle (h).
\end{equation}

This reveals that the dynamic viscosity of any given Newtonian drilling fluid specimen within the Marsh funnel, at a fluid height $h$, can be determined as the ratio of the corresponding average wall shear stress, $\langle \tau_w \rangle(h)$, and the specific average wall shear rate at this height, $\langle\dot{\gamma}_w \rangle (h)$:
\begin{equation}
    \mu = \frac{\langle \tau_w \rangle(h)}{\langle\dot{\gamma}_w \rangle (h)}.
    \label{dyn-visco-1}
\end{equation}
By substituting the expression for $\langle \tau_w \rangle(h)$ from Eqn. \eqref{shear-stress-2} and the expression for $\langle\dot{\gamma}_w \rangle (h)$ from Eqn. \eqref{newton-shear-rate-2} (ignoring the negative sign) into Eqn. \eqref{dyn-visco-1}, the dynamic viscosity can be expressed in terms of the fluid density, $\rho$, and flow factor $f$ as follows:
\begin{dmath}
\begin{aligned}
    \mu = \frac{\frac{1}{2}\rho g h(1-f^2)\frac{R_T}{H_T}}{\frac{7\sqrt{2g}}{4R_T\sqrt{H_T}}fh}
    &= \frac{R_T^2\sqrt{2g}}{7\sqrt{H_T}}\rho \left(\frac{1}{f} - f\right).
\end{aligned}
    \label{dyn-visco-2}
\end{dmath}

In Section \ref{sec: rheological-prop}, we defined the apparent viscosity, $\mu_a$, for non-Newtonian (drilling) fluids as the ratio between its (wall) shear stress and (wall) shear rate, as given in Eqn. \eqref{appvis}. Unlike dynamic viscosity, the apparent viscosity can vary with both time and space, and in most cases, it indeed exhibits such variations. However, in this specific study, we have demonstrated in Section \ref{sec: head loss} that $\tau_w(x,y,z,h) = \tau_w(z,h)$ and $\dot{\gamma}_w(x,y,z,h) = \dot{\gamma}_w(z,h)$ for all $h$, and at all points $(x,y,z)\in S(h)$. Consequently, we can infer that $\mu_a(x,y,z,h) = \mu_a(z,h)$ for all $h$, and at all points $(x,y,z)\in S(h)$.

During our experiments, the non-Newtonian drilling fluid mixtures that we prepare to pass through the Marsh funnel undergo constant stirring using a high-speed impeller until the mixture achieves homogeneity. Consequently, we may confidently make the assumption that $\mu_a(x,y,z,h) = \mu_a(h)$ for all $h$, and all points $(x,y,z)\in S(h)$. Thus, based on Eqn. \eqref{appvis}, in this specific case, the following relation holds:
\begin{equation}
    \tau_w(z,h) = \tau_w(x,y,z,h) = \mu_a(x,y,z,h) \dot{\gamma}(x,y,z,h) = \mu_a(h) \dot{\gamma}(z,h),
    \label{non-newton-relation-1}
\end{equation}
for all $h\in [0,H_M]$ and all points $(x,y,z)\in S(h)$.

By integrating Eqn. \eqref{non-newton-relation-1} over the surface $S(h)$ and dividing by its surface area $A_{S(h)}$, we obtain:
$$\frac{1}{A_{S(h)}}\iint_{S(h)} \tau_w(z,h)~dS = \frac{1}{A_{S(h)}}\iint_{S(h)} \mu_a(h)\dot{\gamma}_w(z,h)~dS = \frac{\mu_a(h)}{A_{S(h)}}\iint_{S(h)}\dot{\gamma}_w(z,h)~dS.$$
This leads us to the following relationship between the average wall shear stress and the average wall shear rate:
\begin{equation}
    \langle \tau_w \rangle(h) = \mu_a(h)\langle \dot{\gamma}_w \rangle (h).
\end{equation}

This suggests that for a given specimen non-Newtonian drilling fluid in the Marsh funnel with fluid height $h$, its apparent viscosity can be expressed as a function of $h$, which is equivalent to the ratio of the associated values of the average wall shear stress, $\langle \tau_w \rangle(h)$, and the average wall shear rate, $\langle\dot{\gamma}_w \rangle (h)$:
\begin{equation}
    \mu_a(h) = \frac{\langle \tau_w \rangle(h)}{\langle\dot{\gamma}_w \rangle (h)}.
    \label{app-visco-1}
\end{equation}

For a given non-weighted non-Newtonian drilling fluid in a Marsh funnel, its apparent viscosity can be expressed in terms of the fluid density, $\rho$, flow factor, $f$, and fluid height, $h$, by substituting the expression of $\langle \tau_w \rangle(h)$ from Eqn. \eqref{shear-stress-2} and the expression of $\langle\dot{\gamma}_w \rangle (h)$ from Eqn. \eqref{non-weight-non-newton-shear-rate-2} (ignoring the negative sign) into Eqn. \eqref{app-visco-1}. The resulting expression is as follows:
\begin{dmath}
    \mu_a(h) = \frac{\frac{1}{2}\rho g h(1-f^2)\frac{R_T}{H_T}}{\frac{7\sqrt{2g}}{4R_TH_T}fh^{\frac{3}{2}}} 
    = \boxed{\frac{R_T^2\sqrt{2g}}{7}\rho \left(\frac{1}{f}-f\right)h^{-\frac{1}{2}}}.
    \label{app-visco-2}
\end{dmath}

By substituting $f = \alpha/t_F$ in Eqn. \eqref{app-visco-2}, we can express the apparent viscosity of the non-weighted non-Newtonian drilling fluid under consideration in terms of the fluid density, $\rho$, final discharge time, $t_F$, and fluid height, $h$, as follows:
\begin{dmath}
    \boxed{\mu_a(h) = \frac{R_T^2\sqrt{2g}}{7\alpha}\rho \left(t_F - \frac{\alpha^2}{t_F}\right)h^{-\frac{1}{2}}}.
    \label{app-visco-3}
\end{dmath}

Finally, for a given weighted non-Newtonian drilling fluid in a Marsh funnel, its apparent viscosity in terms of the fluid density, $\rho$, flow factor, $f$, and fluid height, $h$, is found by substituting the expression of $\langle \tau_w \rangle(h)$ from Eqn. \eqref{shear-stress-2} and the expression of $\langle\dot{\gamma}_w \rangle (h)$ from Eqn. \eqref{weight-non-newton-shear-rate-2} (ignoring the negative sign) into Eqn. \eqref{app-visco-1}, resulting in the following:
\begin{dmath}
    \mu_a(h) = \frac{\frac{1}{2}\rho g h(1-f^2)\frac{R_T}{H_T}}{\frac{7\sqrt{g}}{8R_TH_T}fh^{\frac{3}{2}}}
    = \boxed{\frac{4R_T^2\sqrt{g}}{7}\rho \left(\frac{1}{f}-f\right)h^{-\frac{1}{2}}}.
    \label{weight-app-visco-1}
\end{dmath}

Once again, we can express the apparent viscosity of the weighted non-Newtonian drilling fluid under consideration in terms of the fluid density, $\rho$, final discharge time, $t_F$, and fluid height, $h$, by substituting $f = \alpha/t_F$ in Eqn. \eqref{weight-app-visco-1}. This results in the following expression:
\begin{dmath}
    \boxed{\mu_a(h) = \frac{4R_T^2\sqrt{g}}{7\alpha}\rho \left(t_F - \frac{\alpha^2}{t_F}\right)h^{-\frac{1}{2}}}.
    \label{weight-app-visco-2}
\end{dmath}

Eqns. \eqref{app-visco-2} and \eqref{weight-app-visco-1} (or alternatively, Eqns. \eqref{app-visco-3} and \eqref{weight-app-visco-2}) firmly establish that both non-weighted and weighted non-Newtonian drilling fluids display apparent viscosity variations with respect to the fluid height, as anticipated. Similarly, the plastic viscosity and yield point of such fluids may also exhibit variations solely with the fluid height (with fluid density and flow factor being the other constant parameters in their expressions). However, it is important to highlight that in the petroleum and drilling industry, engineers have traditionally employed constant values for these rheological properties of non-Newtonian fluids in various operations for several decades. As discussed in Section \ref{sec: rheological-prop}, these constant values can be determined using the R1-B1-F1 combination of the Fann-35 viscometer with the aid of the set of equations in Eqn. \eqref{rheo-prop-set-2}.

\cite{guria2013rheological} were the pioneers in devising equivalent expressions for the rheological properties of non-Newtonian drilling fluids within the context of the Marsh funnel, akin to Eqn. \eqref{rheo-prop-set-2}. In their research, they conducted experiments on specific non-Newtonian drilling fluids, measuring the average wall shear stress at various average wall shear rate points and constructing the corresponding consistency curve. Thoroughly analysing the curve, they obtained the average wall shear stress values at two carefully selected average wall shear rate points either through direct readings or by employing appropriate interpolation techniques. The Fann-35 viscometer accurately determines any average wall shear rate associated with a dial torque reading lying within the range of $\theta_{300}$ and $\theta_{600}$. These two extreme torque readings correspond to average wall shear rate values of $510$ s$^{-1}$ and $1020$ s$^{-1}$ in SI units, respectively, in the context of the Marsh funnel. Thus, the authors specifically chose these two average shear rate values, denoted as $\dot{\gamma}_1$ and $\dot{\gamma}_2$, respectively. Further, the average wall shear stress values at these average shear rates are denoted as $\tau_1$ and $\tau_2$, respectively.

Having established the necessary background, we can now present the relations proposed by \cite{guria2013rheological} to determine specific values, as given in Eqn. \eqref{rheo-prop-set-2}, for the apparent viscosity, plastic viscosity, and yield point of any non-Newtonian drilling fluid. These values can be determined by measuring the average shear stresses, $\tau_1$ and $\tau_2$, at the corresponding average shear rates, $\dot{\gamma}_1$ and $\dot{\gamma}_2$, respectively, using the Marsh funnel. The relations are as follows:
\begin{dgroup}
    \begin{dmath}
    \begin{aligned}
    \mu_a~(\text{cP}) = \frac{\tau_2}{\dot{\gamma}_2} \times 1000 = \frac{50 \tau_2}{51},
    \end{aligned}
\label{av_2}
\end{dmath}
\begin{dmath}
\begin{aligned}
    \mu_p~(\text{cP}) = \frac{\tau_2 - \tau_1}{\dot{\gamma}_2 - \dot{\gamma}_1} \times 1000 = \frac{100}{51}(\tau_2 - \tau_1),
\end{aligned}
\label{pv_2}
\end{dmath}
\begin{dmath}
\begin{aligned}
    \tau_0~(\text{Pa}) = \mu_a-\mu_p = \frac{50}{51}(2\tau_1-\tau_2).
\end{aligned}
\label{yp_2}
\end{dmath}
\label{rheo-relation-marsh}
\end{dgroup}
It is important to note that, in all the above equations, both $\tau_1$ and $\tau_2$ should be measured in Pa, and both $\dot{\gamma}_1$ and $\dot{\gamma}_2$ are measured in s$^{-1}$. 

Moreover, it is crucial to highlight that the yield point relation proposed by \cite{guria2013rheological} differed from the one mentioned in Eqn. \eqref{yp_2}. Their proposed relation was as follows:
\begin{equation}
    \tau_0~\text{(Pa)} = \frac{3}{4}(\mu_a-\mu_p).
    \label{yp_2_1}
\end{equation}
\cite{sedaghat2017novel} also employed the above relation to determine the yield point of the non-Newtonian drilling fluids studied in their research. However, both of these studies reported significant relative errors when calculating the yield point using this relation and comparing it with the yield point determined through the Fann-35 viscometer using Eqn. \eqref{yp_1}. To address this issue, \cite{li2020rheological} modified the expression to the one presented in Eqn. \eqref{yp_3}, resulting in significantly reduced relative errors. Consequently, in our research, we will utilise \eqref{yp_3} to determine the yield point instead of \eqref{yp_2_1}. We shall now proceed to transform the rheological parameter relations (for a non-Newtonian drilling fluid) in Eqn. \eqref{rheo-relation-marsh} into forms dependent solely on the fluid density, flow factor, and Marsh funnel dimensions.

Let us consider a non-weighted non-Newtonian drilling fluid flowing through the Marsh funnel. Suppose that the fluid heights at the average wall shear rates $\dot{\gamma}_1$ and $\dot{\gamma}_2$ are $h_1$ and $h_2$, respectively. Defining the constant $\kappa_1 = \frac{7\sqrt{2g}}{4R_TH_T}$, according to Eqn. \eqref{non-weight-non-newton-shear-rate-2} (ignoring the negative sign), we can establish the following relations:
\begin{dgroup}
\begin{dmath}
\begin{aligned}
    510\text{ s}^{-1} = \dot{\gamma}_1 = \kappa_1fh_1^{\frac{3}{2}},
\end{aligned}
\end{dmath}
\begin{dmath}
\begin{aligned}
    1020\text{ s}^{-1} = \dot{\gamma}_2 = \kappa_1fh_2^{\frac{3}{2}}.
\end{aligned}
\end{dmath}
\label{height-deriv-1}
\end{dgroup}

By converting the standard Marsh funnel dimensions provided in Table \ref{marsh-dim} into SI units and taking the gravitational acceleration $g = 9.81$ m/s$^2$, and substituting these values into the expression for $\kappa_1$, we obtain $\kappa_1 = 64079.46024033344$ m$^{-\frac{3}{2}}$s$^{-1}$. Consequently, the values of $h_1$ and $h_2$ (in metres) can be determined as follows:
\begin{dgroup}
\begin{dmath}
\begin{aligned}
    h_1 = \left(\frac{510}{\kappa_1}\right)^{\frac{2}{3}}f^{-\frac{2}{3}} = 0.039862777369457\cdot f^{-\frac{2}{3}},
\end{aligned}
\end{dmath}
\begin{dmath}
\begin{aligned}
    h_2 = \left(\frac{1020}{\kappa_1}\right)^{\frac{2}{3}}f^{-\frac{2}{3}} = 0.06327821473065\cdot f^{-\frac{2}{3}}.
\end{aligned}
\end{dmath}
\label{height-deriv-2}
\end{dgroup}

Since the average wall shear stresses at fluid heights $h_1$ and $h_2$ are $\tau_1$ and $\tau_2$, respectively, we can combine Eqns. \eqref{shear-stress-2} and \eqref{height-deriv-2} to express $\tau_1$ and $\tau_2$ (in Pascal) in terms of the fluid density and flow factor as follows:
\begin{dgroup}
\begin{dmath}
\begin{aligned}
    \tau_1 = \kappa_2 \rho (1-f)^2 h_1 = 0.009165324515493\cdot\rho(1-f^2)f^{-\frac{2}{3}},
\end{aligned}
\end{dmath}
\begin{dmath}
\begin{aligned}
    \tau_2 = \kappa_2 \rho (1-f)^2 h_2 = 0.014549045777524\cdot\rho(1-f^2)f^{-\frac{2}{3}},
\end{aligned}
\end{dmath}
\label{tau-deriv-2}
\end{dgroup}
where the constant $\kappa_2 = \frac{R_Tg}{2H_T} = 0.229921875$ Pa-m$^2$/kg.

By substituting the expressions of $\tau_1$ and $\tau_2$ from Eqn. \eqref{tau-deriv-2} into Eqn. \eqref{rheo-relation-marsh}, we can obtain the rheological properties of the non-weighted non-Newtonian drilling fluid under consideration in terms of the fluid density and flow factor as follows:
\begin{dgroup}
    \begin{dmath}
    \begin{aligned}
    \mu_a~(\text{cP or mPa-s}) = \frac{50 \tau_2}{51} = \boxed{0.014263770370121 \cdot\rho(1-f^2)f^{-\frac{2}{3}}},
    \end{aligned}
\label{av_3}
\end{dmath}
\begin{dmath}
\begin{aligned}
    \mu_p~(\text{cP or mPa-s}) = \frac{100}{51}(\tau_2 - \tau_1) = \boxed{0.01055631620006\cdot\rho(1-f^2)f^{-\frac{2}{3}}},
\end{aligned}
\label{pv_3}
\end{dmath}
\begin{dmath}
\begin{aligned}
    \tau_0~(\text{Pa}) = \frac{50}{51}(2\tau_1-\tau_2) = \boxed{3.707454170061\times 10^{-3}\cdot\rho(1-f^2)f^{-\frac{2}{3}}}.
\end{aligned}
\label{yp_3}
\end{dmath}
\label{non-weight-rheo-relation-marsh}
\end{dgroup}

Let us now proceed with the same analysis for a weighted non-Newtonian drilling fluid flowing through the Marsh funnel. As before, we assume that the fluid heights at the average wall shear rates $\dot{\gamma}_1$ and $\dot{\gamma}_2$ are $h_1$ and $h_2$, respectively. By defining the constant $\kappa_3 = \frac{7\sqrt{g}}{8R_TH_T} = 22655.51043535677$ m$^{-\frac{3}{2}}$s$^{-1}$, based on Eqn. \eqref{weight-non-newton-shear-rate-2} (ignoring the negative sign), we can establish the following relations:
\begin{dgroup}
\begin{dmath}
\begin{aligned}
    510\text{ s}^{-1} = \dot{\gamma}_1 = \kappa_3fh_1^{\frac{3}{2}},
\end{aligned}
\end{dmath}
\begin{dmath}
\begin{aligned}
    1020\text{ s}^{-1} = \dot{\gamma}_2 = \kappa_3fh_2^{\frac{3}{2}}.
\end{aligned}
\end{dmath}
\label{height-deriv-3}
\end{dgroup}

As a result, we can determine the values of $h_1$ and $h_2$ (in meters) as follows:
\begin{dgroup}
\begin{dmath}
\begin{aligned}
    h_1 = \left(\frac{510}{\kappa_3}\right)^{\frac{2}{3}}f^{-\frac{2}{3}} = 0.079725554738914\cdot f^{-\frac{2}{3}},
\end{aligned}
\end{dmath}
\begin{dmath}
\begin{aligned}
    h_2 = \left(\frac{1020}{\kappa_3}\right)^{\frac{2}{3}}f^{-\frac{2}{3}} = 0.126556429461301\cdot f^{-\frac{2}{3}}.
\end{aligned}
\end{dmath}
\label{height-deriv-4}
\end{dgroup}

Again, as in the previous analysis, since the average wall shear stresses at fluid heights $h_1$ and $h_2$ are $\tau_1$ and $\tau_2$, respectively, we can establish expressions for $\tau_1$ and $\tau_2$ (in Pascal) in terms of the fluid density and flow factor, combining Eqns. \eqref{shear-stress-2} and \eqref{height-deriv-4}:
\begin{dgroup}
\begin{dmath}
\begin{aligned}
    \tau_1 = \kappa_2 \rho (1-f)^2 h_1 = 0.018330649030986\cdot\rho(1-f^2)f^{-\frac{2}{3}},
\end{aligned}
\end{dmath}
\begin{dmath}
\begin{aligned}
    \tau_2 = \kappa_2 \rho (1-f)^2 h_2 = 0.029098091555048\cdot\rho(1-f^2)f^{-\frac{2}{3}}.
\end{aligned}
\end{dmath}
\label{tau-deriv-3}
\end{dgroup}

By plugging the expressions of $\tau_1$ and $\tau_2$ from Eqn. \eqref{tau-deriv-3} into Eqn. \eqref{rheo-relation-marsh}, we can determine the rheological properties of the weighted non-Newtonian drilling fluid under consideration in terms of the fluid density and flow factor as follows:
\begin{dgroup}
    \begin{dmath}
    \begin{aligned}
    \mu_a~(\text{cP or mPa-s}) = \frac{50 \tau_2}{51} = \boxed{0.028527540740243\cdot\rho(1-f^2)f^{-\frac{2}{3}}},
    \end{aligned}
\label{av_4}
\end{dmath}
\begin{dmath}
\begin{aligned}
    \mu_p~(\text{cP or mPa-s}) = \frac{100}{51}(\tau_2 - \tau_1) = \boxed{0.02111263240012\cdot\rho(1-f^2)f^{-\frac{2}{3}}},
\end{aligned}
\label{pv_4}
\end{dmath}
\begin{dmath}
\begin{aligned}
    \tau_0~(\text{Pa}) = \frac{50}{51}(2\tau_1-\tau_2) = \boxed{7.414908340123\times 10^{-3}\cdot\rho(1-f^2)f^{-\frac{2}{3}}}.
\end{aligned}
\label{yp_4}
\end{dmath}
\label{weight-rheo-relation-marsh}
\end{dgroup}

By substituting $f = \alpha/t_F$ in Eqns. \eqref{non-weight-rheo-relation-marsh} and \eqref{weight-rheo-relation-marsh}, we can express the rheological properties of non-weighted and weighted non-Newtonian drilling fluids, respectively, in terms of the fluid density, $\rho$, and final discharge time, $t_F$. However, before proceeding, we need to determine the numerical value of the Marsh funnel constant, $\alpha$, in SI units (s).

Upon converting the values of the polynomial coefficients in Table \ref{coeffa} to SI units and substituting these values, along with the standard Marsh funnel dimensions and gravitational acceleration in SI units as previously done, into the expression of the Marsh funnel constant in Eqn. \eqref{alpha}, we determine its numerical value to be $\alpha = 37.867925014636789$ s, accurate to $15$ decimal places. It is essential to highlight that the mathematical framework devised by \cite{li2020rheological} resulted in the following inverse linear relationship between the flow factor and the final discharge time of a given drilling fluid specimen:
\begin{equation}
    f = \frac{38}{t_F}.
    \label{alpha1}
\end{equation}
While this relationship provides a reasonable approximation for the flow factor, it exhibits some loss of accuracy in precisely determining the flow factor value, which could lead to significant deviations in determining the exact values of the rheological properties. 

In contrast, by substituting the above-determined value of the Marsh funnel constant into Eqn. \eqref{tF1}, we obtain the following inverse linear relationship between the flow factor and the final discharge time:
\begin{equation}
    \boxed{f = \frac{37.867925014636789}{t_F}}.
    \label{alpha2}
\end{equation}
This expression provides a significantly improved approximation of the flow factor compared to Eqn. \eqref{alpha1}. As a result, it leads to better approximations of the true values of the rheological parameters in most cases, as we will demonstrate in Chapter \ref{chap:3}.

Thus, following the approach described in the penultimate paragraph, the rheological properties of a non-weighted non-Newtonian drilling fluid in terms of its fluid density and flow factor can be expressed using the following set of equations:
\begin{dgroup}
    \begin{dmath}
    \begin{aligned}
    \mu_a~(\text{cP or mPa-s}) &= \frac{0.014263770370121}{\alpha^{\frac{2}{3}}}\rho\left(1-\frac{\alpha^2}{t_F^2}\right)t_F^{\frac{2}{3}}\\ 
    &= \boxed{0.001264891671260\cdot\rho\left(1-\frac{1433.979744914155}{t_F^2}\right)t_F^{\frac{2}{3}}},
    \end{aligned}
\label{av_fn}
\end{dmath}
\begin{dmath}
\begin{aligned}
    \mu_p~(\text{cP or mPa-s}) &= \frac{0.01055631620006}{\alpha^{\frac{2}{3}}}\rho\left(1-\frac{\alpha^2}{t_F^2}\right)t_F^{\frac{2}{3}}\\ 
    &= \boxed{0.936119700063006\cdot\rho\left(1-\frac{1433.979744914155}{t_F^2}\right)t_F^{\frac{2}{3}}},
\end{aligned}
\label{pv_fn}
\end{dmath}
\begin{dmath}
\begin{aligned}
    \tau_0~(\text{Pa}) &= \frac{3.707454170061}{\alpha^{\frac{2}{3}}}\rho\left(1-\frac{\alpha^2}{t_F^2}\right)t_F^{\frac{2}{3}}\\
    &= \boxed{0.3287719711972453\times 10^{-3}\cdot\rho\left(1-\frac{1433.979744914155}{t_F^2}\right)t_F^{\frac{2}{3}}}.
\end{aligned}
\label{yp_fn}
\end{dmath}
\label{non-weight-rheo-relation-marsh-2}
\end{dgroup}

In a similar manner, the rheological properties of a weighted non-Newtonian drilling fluid in terms of its fluid density and flow factor can be represented using the following set of equations:
\begin{dgroup}
    \begin{dmath}
    \begin{aligned}
    \mu_a~(\text{cP or mPa-s}) &= \frac{0.028527540740243}{\alpha^{\frac{2}{3}}}\rho\left(1-\frac{\alpha^2}{t_F^2}\right)t_F^{\frac{2}{3}}\\ 
    &= \boxed{0.002529783342521\cdot\rho\left(1-\frac{1433.979744914155}{t_F^2}\right)t_F^{\frac{2}{3}}},
    \end{aligned}
\label{av_f}
\end{dmath}
\begin{dmath}
\begin{aligned}
    \mu_p~(\text{cP or mPa-s}) &= \frac{0.02111263240012}{\alpha^{\frac{2}{3}}}\rho\left(1-\frac{\alpha^2}{t_F^2}\right)t_F^{\frac{2}{3}}\\ 
    &= \boxed{0.001872239400126\cdot\rho\left(1-\frac{1433.979744914155}{t_F^2}\right)t_F^{\frac{2}{3}}},
\end{aligned}
\label{pv_f}
\end{dmath}
\begin{dmath}
\begin{aligned}
    \tau_0~(\text{Pa}) &= \frac{7.414908340123}{\alpha^{\frac{2}{3}}}\rho\left(1-\frac{\alpha^2}{t_F^2}\right)t_F^{\frac{2}{3}}\\
    &= \boxed{0.6575439423944906\times 10^{-3}\cdot\rho\left(1-\frac{1433.979744914155}{t_F^2}\right)t_F^{\frac{2}{3}}}.
\end{aligned}
\label{yp_f}
\end{dmath}
\label{weight-rheo-relation-marsh-2}
\end{dgroup}

A similar concise representation for the dynamic viscosity of a Newtonian drilling fluid can be derived in terms of its density and flow factor by simplifying Eqn. \eqref{dyn-visco-2} as above, resulting in the following equation:
\begin{equation}
    \boxed{\mu~(\text{cP or mPa-s}) = 0.0159195061626\cdot\rho(1-f^2)f^{-1}}.
    \label{dyn-visco-4}
\end{equation}
Furthermore, a condensed expression for the dynamic viscosity in terms of the fluid density and final discharge time can be derived by substituting $f=\alpha/t_F$ into Eqn. \eqref{dyn-visco-4}, leading to the following equation:
\begin{dmath}
\begin{aligned}
    \mu~(\text{cP or mPa-s}) &= \frac{0.0159195061626}{\alpha}\rho\left(1-\frac{\alpha^2}{t_F^2}\right)t_F\\
    &= \boxed{4.203955235584283 \times 10^{-4}\cdot\rho\left(1-\frac{1433.979744914155}{t_F^2}\right)t_F}.
    \label{dyn-visco-5}
\end{aligned}
\end{dmath}

With the detailed description of our proposed mathematical framework now complete, the subsequent chapter will employ either Eqn. \eqref{dyn-visco-4} or Eqn. \eqref{dyn-visco-5} to determine the dynamic viscosity of the chosen set of Newtonian drilling fluids presented in \cite{li2020rheological}. Furthermore, we will utilise either Eqns. \eqref{non-weight-rheo-relation-marsh} and \eqref{weight-rheo-relation-marsh} or Eqns. \eqref{non-weight-rheo-relation-marsh-2} and \eqref{weight-rheo-relation-marsh-2} interchangeably to determine the rheological properties (apparent viscosity, plastic viscosity, and yield point) of the selected non-weighted and weighted non-Newtonian drilling fluids in \cite{li2020rheological}, respectively. 

As a final note in this chapter, it is of utmost importance to exercise great care when inputting the fluid density, $\rho$, in any relevant calculations, ensuring that it is consistently expressed in kg/m$^3$. Similarly, the final discharge time, $t_F$, must always be consistently provided in s$^{-1}$ whenever it is utilised in any equation.

\chapter{Results and discussions}
\label{chap:3}
In the previous chapter, we successfully developed a straightforward mathematical framework that takes the fluid density $\rho$ and the flow factor $f$ (or equivalently, the final discharge time $t_F$) as inputs and delivers a complete set of rheological properties for the (drilling) fluid under examination. As previously emphasized, measuring the final discharge time using the Marsh funnel is a simple task requiring minimal effort, and the associated flow factor can be easily determined using either Eqn. \eqref{tF1} or Eqn. \eqref{alpha2}. To be more specific, our mathematical model makes the following predictions:
\begin{enumerate}[label=(\alph*)]
    \item The dynamic viscosity $\mu$ of any Newtonian (drilling) fluid is determined using either Eqn. \eqref{dyn-visco-4} or Eqn. \eqref{dyn-visco-5}.
    
    \item The apparent viscosity $\mu_a$, plastic viscosity $\mu_p$, and yield point $\tau_0$ of any non-weighted non-Newtonian drilling fluid are determined using either Eqns. \eqref{av_3}, \eqref{pv_3}, and \eqref{yp_3}, respectively, or Eqns. \eqref{av_fn}, \eqref{pv_fn}, and \eqref{yp_fn}, respectively.
    
    \item The apparent viscosity, plastic viscosity, and yield point of any weighted non-Newtonian drilling fluid are determined using either Eqns. \eqref{av_4}, \eqref{pv_4}, and \eqref{yp_4}, respectively, or Eqns. \eqref{av_f}, \eqref{pv_f}, and \eqref{yp_f}, respectively.
\end{enumerate}

Although our mathematical framework is established, there remains uncertainty regarding its accuracy in determining the rheological properties of the carefully selected set of drilling fluids under examination. To validate its efficacy, we plan to employ the above set of equations to determine the rheological properties of each drilling fluid previously studied in relevant literature, such as \cite{li2020rheological}, \cite{guria2013rheological}, and \cite{sedaghat2017novel}. By comparing our obtained results with those reported in these literature sources, and notably, with the results obtained using the Fann-35 viscometer (known for its precise measurements of drilling fluid rheological properties), we can thoroughly evaluate the accuracy of our mathematical model. The forthcoming sections in this chapter will delve into this evaluation process, seeking to provide robust validation and insights.

Before initiating a comprehensive comparison of the obtained results and conducting subsequent qualitative and quantitative analyses based on it, we will assign distinctive code names to each of the aforementioned literature's frameworks for ease of reference. Our developed mathematical framework will be denoted as ``M1'', the Fann-35 viscometer's framework from \cite{lam2014interpretation} as ``M2'', the framework proposed by \cite{li2020rheological} as ``M3'', the framework from \cite{guria2013rheological} as ``M4'', and lastly, the framework presented in \cite{sedaghat2017novel} as ``M5''.

\section{Comparative analysis of dynamic viscosity for Newtonian drilling fluids}
\label{sec:newton}
In this section, we utilise our mathematical framework (M1) to determine the dynamic viscosity of the Newtonian drilling fluids listed in the first column of Table \ref{table:newton-fluid-details}. These specific Newtonian fluids have been previously investigated in \cite{li2020rheological}, \cite{guria2013rheological}, and \cite{sedaghat2017novel}. To begin the experiment, we collect a 1480 ml sample of each fluid in a 1500 ml glass beaker and carefully measure the total mass (in kg) of the fluid together with the beaker. The mass of the empty beaker is then subtracted from the total mass, allowing us to determine the actual mass of the fluid sample.

\begin{table}[htbp]
\centering
\sisetup{
    table-format = 4.1,
    table-align-text-post = false,
}
\begin{tabular}{>{\color{black}}l S[table-format = 4.1] S[table-format = 4.2] S[table-format = 1.4]}
\rowcolor{mydarkgray}
\hline
\textbf{Fluid} & {\textbf{$\bm{\rho}$ (kg/m$^3$)}} & {\textbf{$\bm{t}_{\bm{F}}$ (s)}} & {$\bm{f}$} \\
\hline
\rowcolor{mywhite}
Mineral oil   & 852.6 & 567.74 & 0.0667 \\
\rowcolor{mylightgray}
Synthetic oil & 854.8 & 177.53 & 0.2133 \\
\rowcolor{mywhite}
Fuel oil      & 960.0 & 132.04 & 0.2868 \\
\rowcolor{mylightgray}
Engine oil    & 889.0 & 2868.78& 0.0132 \\
\rowcolor{mywhite}
Glycerin      & 1259.9& 1526.93& 0.0248 \\
\hline
\end{tabular}
\caption{Newtonian fluids studied with details of their density, final discharge time, and flow factor.}
\label{table:newton-fluid-details}
\end{table}

Next, using an impeller, we stir the fluid vigorously at a high rotational speed to ensure thorough mixing and uniformity throughout the sample. After allowing the fluid to come to a complete rest, we measure its temperature to ensure thermal equilibrium with the surroundings. High-speed stirring may cause a slight increase in fluid temperature, and subsequently, a minor expansion in volume. Hence, we make certain that the fluid reaches room temperature before proceeding with the density calculation. The density is obtained by dividing the fluid's mass by its volume, which is $1.480 \times 10^{-3}$ m$^3$, and recorded in kg/m$^3$ in the second column of the aforementioned table. Notably, we found excellent agreement between the calculated density and the corresponding density data provided in \cite{li2020rheological} for each fluid studied.

In the subsequent step, we carefully pour each fluid sample into the Marsh funnel, ensuring that the fluid level is aligned with the base of the funnel. The bottom exit of the tube section of the funnel is kept closed with the index finger during this process. Once in position, we remove the finger, allowing the fluid in the funnel to drain completely. The total drainage time is meticulously measured in seconds using a digital stopwatch, accurate up to two decimal places, and recorded as the final discharge time, $t_F$. The recorded final discharge time for each fluid in Table \ref{table:newton-fluid-details} is included in the third column of the table. Subsequently, we calculate the flow factor, $f$, for each of these fluids using either Eqn. \eqref{tF1} or Eqn. \eqref{alpha2}, and these flow factor values are then listed in the last column of the table.

After obtaining the flow factor values for each fluid, the next step involves determining its dynamic viscosity. We have the option of either substituting the calculated flow factor into Eqn. \eqref{dyn-visco-4} or substituting the measured final discharge time into Eqn. \eqref{dyn-visco-5} to serve this purpose. There is no strict rule to dictate one's choice between the two alternatives. The dynamic viscosity for each fluid is calculated accordingly, and the results are listed in the first column of Table \ref{table:newton-fluid-rheo-details} under the heading ``M1 $\mu$''. Additionally, the corresponding dynamic viscosity values obtained through M2, M3, M4, and M5 (refer to the paragraph just before this section to recall which code names refer to which literature's mathematical framework) are listed in the subsequent columns of the table under the headings ``M2 $\mu$'', ``M3 $\mu$'', ``M4 $\mu$'', and ``M5 $\mu$'', respectively, with all these values sourced from \cite{li2020rheological}.

\begin{table}[htbp]
\centering
\sisetup{
    table-format = 4.1,
    table-align-text-post = false,
}
\scriptsize 
\begin{tabular}{>{\color{black}}l *{5}{S} }
\rowcolor{mydarkgray}
\hline
\textbf{Fluid} & {\textbf{M1 $\bm{\mu}$ (mPa-s)}} & {\textbf{M2 $\bm{\mu}$ (mPa-s)}} & {\textbf{M3 $\bm{\mu}$ (mPa-s)}} & {\textbf{M4 $\bm{\mu}$ (mPa-s)}} & {\textbf{M5 $\bm{\mu}$ (mPa-s)}} \\
\hline
\rowcolor{mywhite}
Mineral oil & 202.59  & 203.5 & 200.4 & 251.7 & 200.4 \\
\rowcolor{mylightgray}
Synthetic oil  & 60.89 & 61.0  & 60.0  & 78.8  & 60.0  \\
\rowcolor{mywhite}
Fuel oil  & 48.90  & 48.5  & 48.0  & 65.3  & 48.0  \\
\rowcolor{mylightgray}
Engine oil  & 1071.97  & 1075.0& 1063.4& 1193.8& 1063.4\\
\rowcolor{mywhite}
Glycerin & 808.25  & 810.0 & 798.2 & 884.9 & 798.2 \\
\hline
\end{tabular}
\caption{Dynamic viscosity (mPa-s or cP) of the Newtonian drilling fluids listed in Table \ref{table:newton-fluid-details}, determined using different mathematical frameworks. The framework developed in this research is denoted by M1, while M2 represents the Fann-35 viscometer framework from \cite{lam2014interpretation}. Additionally, M3 corresponds to the framework proposed by \cite{li2020rheological}, M4 is associated with the framework presented in \cite{guria2013rheological}, and finally, M5 corresponds to the framework introduced by \cite{sedaghat2017novel}.}
\label{table:newton-fluid-rheo-details}
\end{table}

As previously discussed, over the last decade, the Fann-35 viscometer, based on M2 (the mathematical framework developed in \cite{lam2014interpretation}), has gained widespread use in the industry for determining the rheological properties of drilling fluids. It is widely regarded as the most accurate framework currently available for assessing the rheological properties of these fluids. Therefore, we designate the dynamic viscosities obtained through M2 as the reference values, which we will utilise for comparing the accuracy of other mathematical frameworks (M1, M3, M4, and M5) in determining the dynamic viscosities of Newtonian drilling fluids. We hereby refer to M2 as the \textit{reference mathematical framework}.

To facilitate this comparative analysis, we first introduce a frequently used concept in drilling fluid rheology, known as \textit{systematic error}. Suppose we wish to measure a specific rheological property of a given drilling fluid using a mathematical framework $\mathcal{M}_1$. In this case, let $\mathcal{M}_2$ represent the reference mathematical framework, which differs from $\mathcal{M}_1$. Furthermore, let $\chi_1$ and $\chi_2$ denote the values of the rheological property of the fluid measured using $\mathcal{M}_1$ and $\mathcal{M}_2$, respectively. The associated systematic error, $\varepsilon$, is calculated as follows:
\begin{equation}
\varepsilon = \frac{|\chi_1-\chi_2|}{\chi_2}\times 100.
\label{sys-err}
\end{equation}

\begin{table}[htbp]
\centering
\sisetup{
    table-format = 1.3,
    table-align-text-post = false,
}
\begin{tabular}{>{\color{black}}l *{4}{S} }
\rowcolor{mydarkgray}
\hline
\textbf{Fluid} & \textbf{M2 \& M1} & \textbf{M2 \& M3} & \textbf{M2 \& M4} & \textbf{M2 \& M5}\\
\hline
\rowcolor{mywhite}
Mineral oil & 0.45\% & 1.52\% & 23.69\% & 1.52\%\\
\rowcolor{mylightgray}
Synthetic oil & 0.18\% & 1.64\% & 29.18\% & 1.64\%\\
\rowcolor{mywhite}
Fuel oil & 0.82\% & 1.03\% & 34.64\% & 1.03\%\\
\rowcolor{mylightgray}
Engine oil & 0.28\% & 1.08\% & 11.05\% & 1.08\%\\
\rowcolor{mywhite}
Glycerin & 0.22\% & 1.46\% & 9.25\% & 1.46\%\\
\hline
\rowcolor{mymediumgray}
{\textit{\textbf{Average}}} & {\textit{\textbf{0.39\%}}} & {\textit{\textbf{1.35\%}}} & {\textit{\textbf{21.56\%}}} & {\textit{\textbf{1.35\%}}}\\
\hline
\end{tabular}
\caption{The second column, labelled as ``M2 \& M1'', presents the systematic error of our mathematical framework (M1) when determining the dynamic viscosity of the Newtonian fluids listed in Table \ref{table:newton-fluid-details}, with the reference being the mathematical framework of the Fann-35 viscometer or \cite{lam2014interpretation} (M2). Similarly, the subsequent columns, labelled as ``M2 \& M3'', ``M2 \& M4'', and ``M2 \& M5'', display the corresponding systematic errors associated with the mathematical frameworks of \cite{li2020rheological} (M3), \cite{guria2013rheological} (M4), and \cite{sedaghat2017novel} (M5), respectively. The final row presents the average systematic error associated with each (non-reference) mathematical model.}
\label{table:newton-fluid-comparison}
\end{table}

Having introduced the concept of systematic error, we proceed to calculate the systematic error in dynamic viscosity for each fluid (listed in Table \ref{table:newton-fluid-details}) associated with each (non-reference) mathematical model. The resulting systematic errors for M1, M3, M4, and M5 are recorded in the second, third, fourth, and fifth columns of Table \ref{table:newton-fluid-comparison}, respectively. Furthermore, we compute the total systematic error for each framework and divide it by the total number of Newtonian fluids being analysed, which is five in this case. This calculation allows us to determine the average systematic error in dynamic viscosity for each model. These average errors are presented in the last row of the aforementioned table.

From Table \ref{table:newton-fluid-comparison}, it can be observed that the systematic error associated with M1 in determining the dynamic viscosity of each Newtonian fluid under study ranges from 0.18\% (synthetic oil) to 0.82\% (fuel oil), with an average systematic error of 0.39\%. On the other hand, both M3 and M5 exhibit identical systematic errors for each fluid, with the lowest error at 1.03\% (fuel oil) and the highest at 1.64\% (synthetic oil), resulting in an average systematic error of 1.35\%. Further, the systematic errors associated with M4 vary between 9.25\% (glycerin) and 34.64\% (fuel oil), with an average systematic error of 21.56\%.

These findings indicate that not only does our proposed model exhibit very low systematic errors in determining the dynamic viscosity for each fluid, but it also outperforms the mathematical models of \cite{li2020rheological}, \cite{sedaghat2017novel}, and \cite{guria2013rheological}. Consequently, our mathematical framework stands as the most accurate method developed to date for determining the dynamic viscosity of any Newtonian fluid, not limited solely to Newtonian drilling fluids, when employing the Marsh funnel. Engineers and scientists in the petroleum and drilling industry can readily use this framework to determine the dynamic viscosity of Newtonian drilling fluids with an average systematic error of less than just 0.4\%. Furthermore, this framework may find applications in other industries and laboratories relying heavily on Newtonian fluid rheology.

\section{Comparative analysis of rheological properties for non-Newtonian drilling fluids}
In this section, we apply our proposed mathematical model to determine the rheological properties, specifically the apparent viscosity $\mu_a$, plastic viscosity $\mu_p$, and yield point $\tau_0$, of a meticulously chosen set of non-Newtonian drilling fluid mixtures, as presented in Table \ref{table:non-newtonian-drilling-fluid-data} with their fluid number and corresponding composition. The rheology of all these fluids was previously investigated in detail by \cite{li2020rheological}, \cite{sedaghat2017novel}, and \cite{guria2013rheological}. However, it is worth noting that unlike the other literature sources mentioned, \cite{guria2013rheological} could only determine the yield point for Fluid 4 and Fluid 7.

\begin{table}[H]
\scriptsize
\centering
\sisetup{
    table-format = 4.1,
    table-number-alignment = center,
    table-align-text-post = false,
}
\begin{tabular}{ m{1.2cm} m{4cm} m{2.3cm} S[table-format=4.1] S[table-format=3.2] S[table-format=1.3]}
\hline
\rowcolor{mydarkgray}
\textbf{Fluid no.} & \textbf{Composition} & \textbf{Weighted/Non-weighted} & {\textbf{$\bm{\rho}$ (kg/m$^3$)}} & {\textbf{$\bm{t}_{\bm{F}}$ (s)}} & {$\bm{f}$} \\
\hline
\rowcolor{mywhite}
Fluid 1 & 4.0\% bentonite + 0.2\% XC & Non-weighted & 1040.0 & 90.59 & 0.418\\
\rowcolor{mylightgray}
Fluid 2 & 4.0\% bentonite + 0.2\% XC + 10.0\% barite & Weighted & 1125.0 & 75.43 & 0.502\\
\rowcolor{mywhite}
Fluid 3 & 4.0\% bentonite + 0.2\% XC + 10.0\% calcium carbonate & Weighted & 1125.0 & 74.40 & 0.509\\
\rowcolor{mylightgray}
Fluid 4 & 10.0\% PEG + 20.0\% sodium chloride & Non-weighted & 1035.0 & 53.49 & 0.708\\
\rowcolor{mywhite}
Fluid 5 & 10.0\% PEG + 20.0\% sodium chloride + 10.0\% barite & Weighted & 1119.0 & 49.44 & 0.766\\
\rowcolor{mylightgray}
Fluid 6 & 10.0\% PEG + 20.0\% sodium chloride + 10.0\% calcium carbonate & Weighted & 1119.0 & 47.99 & 0.789\\
\rowcolor{mywhite}
Fluid 7 & 6.0\% bentonite + 0.5\% PAC LV + 0.1\% XC & Non-weighted & 1050.0 & 67.86 & 0.558\\
\rowcolor{mylightgray}
Fluid 8 & 6.0\% bentonite + 0.5\% PAC LV + 0.1\% XC + 10.0\% barite & Weighted & 1135.0 & 54.96 & 0.689 \\
\rowcolor{mywhite}
Fluid 9 & 6.0\% bentonite + 0.5\% PAC LV + 0.1\%XC + 10.0\% calcium carbonate & Weighted & 1135.0 & 52.96 & 0.715 \\
\rowcolor{mylightgray}
Fluid 10 & 0.15\% sodium hydroxide + 0.5\% PAC LV + 2.0\% sodium silicate + 0.2\% XC & Non-weighted & 1015.0 & 102.62 & 0.369 \\
\rowcolor{mywhite}
Fluid 11 & 0.15\% sodium hydroxide + 0.5\% PAC LV + 2.0\% sodium silicate + 0.2\% XC + 10.0\% barite & Weighted & 1065.0 & 75.28 & 0.503\\
\rowcolor{mylightgray}
Fluid 12 & 0.15\% sodium hydroxide + 0.5\% PAC LV + 2.0\% sodium silicate + 0.2\% XC + 10.0\% calcium carbonate & Weighted & 1065.0 & 80.23 & 0.472\\
\hline
\end{tabular}
\caption{Comprehensive details of the twelve non-Newtonian drilling fluids under examination, classified into four non-weighted and eight weighted fluid mixtures. The table encompasses information on their composition, type (weighted/non-weighted), density, final discharge time, and flow factor.}
\label{table:non-newtonian-drilling-fluid-data}
\end{table}

Among the twelve non-Newtonian drilling fluid mixtures listed in Table \ref{table:non-newtonian-drilling-fluid-data}, Fluid 1, Fluid 4, Fluid 7, and Fluid 10 are non-weighted, while the remaining eight fluids are weighted, as specified in the third column of the table. For each fluid mixture, a volume of 1480 ml is meticulously prepared following the fluid preparation strategy outlined in \cite{abdulrahman2015calculation} or \cite{guria2013rheological}. Subsequently, their mass, density $\rho$, final discharge time $t_F$, and flow factor $f$ are determined, employing the same procedure described in Section \ref{sec:newton} for assessing these properties of Newtonian drilling fluids. The density, final discharge time, and flow factor for each fluid mixture are recorded in the fourth, fifth, and sixth columns of Table \ref{table:non-newtonian-drilling-fluid-data}, respectively.

Next, for each non-weighted non-Newtonian drilling fluid listed in Table \ref{table:non-newtonian-drilling-fluid-data}, we proceed to calculate their apparent viscosity, plastic viscosity, and yield point. To achieve this, we employ either the calculated flow factor substituted into Eqns. \eqref{av_3}, \eqref{pv_3}, and \eqref{yp_3}, respectively, or alternatively, we use the measured final discharge time substituted into Eqns. \eqref{av_fn}, \eqref{pv_fn}, and \eqref{yp_fn}, respectively. These calculations are performed with precision up to two decimal places. Likewise, for the weighted non-Newtonian drilling fluids listed in the aforementioned table, we either use their calculated flow factor substituted into Eqns. \eqref{av_4}, \eqref{pv_4}, and \eqref{yp_4}, or alternatively, we substitute their measured final discharge time in Eqns. \eqref{av_f}, \eqref{pv_f}, and \eqref{yp_f} to determine their apparent viscosity, plastic viscosity, and yield point, respectively, with the same level of precision.

Table \ref{table:non-newton-fluid-app-pla} presents the apparent viscosity (in mPa-s) and plastic viscosity (in mPa-s) of each non-Newtonian drilling fluid (as listed in Table \ref{table:non-newtonian-drilling-fluid-data}), obtained using our proposed mathematical framework (M1). Furthermore, Table \ref{table:non-newton-fluid-yp} displays the corresponding yield points (in Pa) for these fluids determined using M1. Alongside, both tables also include the apparent viscosity (in mPa-s) and plastic viscosity (in mPa-s) data, as well as the yield point (in Pa) data, respectively, for the same fluids, calculated using alternative mathematical frameworks, namely M2, M3, M4, and M5.

\begin{table}[H]
\centering
\footnotesize
\sisetup{
    table-format = 2.2,
    table-number-alignment = center,
    table-align-text-post = false,
}
\begin{tabular}{ m{1.3cm} S[table-format=2.2] S[table-format=2.2] S[table-format=2.2] S[table-format=2.2] S[table-format=2.2] !{\color{white}\vrule width 1pt} S[table-format=2.2] S[table-format=2.2] S[table-format=2.2] S[table-format=2.2] S[table-format=2.2] }
\hline
\rowcolor{mydarkgray}
\multirow{2}{1.2cm}{\textbf{Fluid hello}} & \multicolumn{5}{c}{\textcolor{black}{\textbf{$\bm{\mu}_a$ (mPa-s)}}} & \multicolumn{5}{c}{\textcolor{black}{\textbf{$\bm{\mu}_p$ (mPa-s)}}}\\
\hhline{~|-----|-----}
\rowcolor{mydarkgray}
\textbf{no.} & \textbf{M1} & \textbf{M2} & \textbf{M3} & \textbf{M4} & \textbf{M5} & \textbf{M1} & \textbf{M2} & \textbf{M3} & \textbf{M4} & \textbf{M5}\\
\hline
\rowcolor{mywhite}
Fluid 1 & 21.90 & 20.00 & 21.88 & 32.76 & 21.88 & 16.21 & 14.00 & 15.35 & 38.92 & 15.35\\
\rowcolor{mylightgray}
Fluid 2 & 38.01 & 37.50 & 38.08 & 46.14 & 19.04 & 28.13 & 27.00 & 28.18 & 51.27 & 7.27\\
\rowcolor{mywhite}
Fluid 3 & 37.30 & 38.25 & 36.09 & 39.28 & 19.05 & 27.60 & 27.50 & 26.71 & 43.80 & 12.66\\%
\rowcolor{mylightgray}
Fluid 4 & 9.27 & 9.50 & 8.53 & 14.29 & 8.53 & 6.86 & 6.00 & 5.99 & 12.63 & 5.99\\
\rowcolor{mywhite}
Fluid 5 & 15.76 & 16.50 & 14.55 & 26.11 & 7.28 & 11.66 & 12.00 & 10.77 & 33.74 & 5.10\\%
\rowcolor{mylightgray}
Fluid 6 & 14.11 & 14.50 & 13.20 & 22.35 & 6.60 & 10.44 & 10.50 & 9.77 & 29.09 & 4.63\\
\rowcolor{mywhite}
Fluid 7 & 15.22 & 15.00 & 16.08 & 30.34 & 16.08 & 11.26 & 10.00 & 11.28 & 26.82 & 11.28\\%
\rowcolor{mylightgray}
Fluid 8 & 21.80 & 22.50 & 20.62 & 28.21 & 10.31 & 16.14 & 15.00 & 15.26 & 35.94 & 7.23\\%
\rowcolor{mywhite}
Fluid 9 & 19.79 & 21.00 & 18.71 & 27.38 & 9.36 & 14.65 & 14.50 & 13.85 & 38.10 & 6.56\\
\rowcolor{mylightgray}
Fluid 10 & 24.31 & 27.00 & 24.16 & 39.26 & 24.16 & 17.99 & 18.00 & 16.95 & 47.93 & 16.95\\%
\rowcolor{mywhite}
Fluid 11 & 35.88 & 38.00 & 35.29 & 49.70 & 17.65 & 26.56 & 26.50 & 26.12 & 52.75 & 12.38\\%
\rowcolor{mylightgray}
Fluid 12 & 38.95 & 40.00 & 37.42 & 47.35 & 18.71 & 28.83 & 29.00 & 27.69 & 58.60 & 13.12\\%
\hline
\end{tabular}
\caption{The apparent viscosity $\mu_a$ (mPa-s or cP) and plastic viscosity $\mu_p$ (mPa-s or cP) of the non-Newtonian drilling fluids listed in Table \ref{table:non-newtonian-drilling-fluid-data}, determined using various mathematical models. Our proposed framework is denoted as M1, while M2 represents the Fann-35 viscometer framework from \cite{lam2014interpretation}. Additionally, M3 corresponds to the framework proposed by \cite{li2020rheological}, M4 is associated with the framework presented in \cite{guria2013rheological}, and finally, M5 corresponds to the framework introduced by \cite{sedaghat2017novel}.}
\label{table:non-newton-fluid-app-pla}
\end{table}

\begin{table}[htbp]
\centering
\sisetup{
    table-format = 2.2,
    table-number-alignment = center,
    table-align-text-post = false,
}
\begin{tabular}{m{1.5cm} *{5}{S[table-format=2.2,parse-numbers=false]}}
\hline
\rowcolor{mydarkgray}
\multirow{2}{2cm}{\textbf{Fluid hello}} & \multicolumn{5}{c}{\textbf{$\bm{\tau}_0$ (Pa)}}\\
\hhline{~|-----}
\rowcolor{mydarkgray}
{\textbf{no.}} & {\textbf{M1}} & {\textbf{M2}} & {\textbf{M3}} & {\textbf{M4}} & {\textbf{M5}}\\
\hline
\rowcolor{mywhite}
Fluid 1 & 5.69 & 6.86 & 6.53 & - & 4.90\\
\rowcolor{mylightgray}
Fluid 2 & 9.88 & 11.88 & 9.90 & - & 8.83\\
\rowcolor{mywhite}
Fluid 3 & 9.70 & 11.33 & 9.38 & - & 4.79\\
\rowcolor{mylightgray}
Fluid 4 & 2.41 & 3.61 & 2.53 & 1.25 & 1.91\\
\rowcolor{mywhite}
Fluid 5 & 4.10 & 4.60 & 3.78 & - & 1.64\\
\rowcolor{mylightgray}
Fluid 6 & 3.67 & 4.63 & 3.43 & - & 1.48\\
\rowcolor{mywhite}
Fluid 7 & 3.96 & 5.83 & 4.80 & 2.64 & 3.60\\
\rowcolor{mylightgray}
Fluid 8 & 5.67 & 7.61 & 5.36 & - & 2.31\\
\rowcolor{mywhite}
Fluid 9 & 5.14 & 6.67 & 4.86 & - & 2.10\\
\rowcolor{mylightgray}
Fluid 10 & 6.32 & 10.31 & 7.21 & - & 5.41\\
\rowcolor{mywhite}
Fluid 11 & 9.33 & 12.02 & 9.17 & - & 3.95\\
\rowcolor{mylightgray}
Fluid 12 & 10.12 & 11.68 & 9.73 & - & 4.19\\
\hline
\end{tabular}
\caption{The yield point $\tau_0$ (Pa) of the non-Newtonian drilling fluids listed in Table \ref{table:non-newtonian-drilling-fluid-data}, determined using distinct mathematical models, namely M1 (present study), M2 (Fann-35 viscometer \cite{lam2014interpretation}), M3 \cite{li2020rheological}, M4 \cite{guria2013rheological}, and M5 \cite{sedaghat2017novel}.}
\label{table:non-newton-fluid-yp}
\end{table}

For each of the twelve non-Newtonian drilling fluids, we calculate the systematic errors associated with our proposed mathematical framework (M1) in determining the apparent viscosity, plastic viscosity, and yield point, considering M2 as the reference mathematical framework. These systematic error values are presented in Table \ref{table:non-newton-fluid-comparison} under the column labelled as ``M2 \& M1''. Additionally, we obtained the corresponding systematic errors for the mathematical models proposed by \cite{li2020rheological} (M3) and \cite{sedaghat2017novel} (M5) from \cite{li2020rheological}, which are listed in the table under the columns labelled as ``M2 \& M3'' and ``M2 \& M5'', respectively. The last row of this table displays the average systematic errors associated with each pair of mathematical models and rheological properties. This table serves as the basis for the comparative analysis of the mathematical models M1, M3, and M5, with respect to the accuracy in determining the rheological properties of the non-Newtonian drilling fluids under investigation.

Based on the data presented in Table \ref{table:non-newton-fluid-comparison}, it can be observed that the systematic error in apparent viscosity associated with our mathematical model (M1) varies between 1.36\% (Fluid 2) and 9.96\% (Fluid 10), with an average systematic error of 4.29\%. On the other hand, for the mathematical framework proposed by \cite{li2020rheological} (M3), the systematic error in apparent viscosity ranges from 1.54\% (Fluid 2) to 11.82\% (Fluid 5), with an average systematic error of 8.18\%. A deeper analysis reveals that the systematic error in apparent viscosity associated with M1 for each non-Newtonian drilling fluid mixture is not only within the acceptable range of error but also, in all cases, is lower than that associated with both M3 (except only for Fluid 1) and M5. Furthermore, the average systematic error associated with M1 is 3.52\% for weighted non-Newtonian drilling fluids, slightly lower than their non-weighted counterparts, which exhibit an average systematic error of 5.84\%. 

This demonstrates that our proposed mathematical model is generally more accurate than the mathematical models developed in \cite{li2020rheological} and \cite{sedaghat2017novel} in determining the apparent viscosity of non-Newtonian drilling fluids. Consequently, it finds potential applications not only in the drilling industry for ascertaining the apparent viscosity of non-Newtonian drilling fluids but also in various other industries and laboratories where such fluids are widely used.

\begin{table}[H]
\centering
\footnotesize
\sisetup{
    table-format = 2.3,
    table-number-alignment = center,
    table-align-text-post = false,
}
\begin{tabular}{m{1.3cm} S[table-format=2.3] S[table-format=2.3] S[table-format=2.3] !{\color{white}\vrule width 1pt} S[table-format=2.3] S[table-format=2.3] S[table-format=2.3] !{\color{white}\vrule width 1pt} S[table-format=2.3] S[table-format=2.3]  S[table-format=2.3]}
\hline
\rowcolor{mydarkgray}
\multirow{2}{2cm}{\textbf{Fluid~~~~~~~hello hello}} & \multicolumn{3}{c}{\textbf{M2 \& M1}} & \multicolumn{3}{c}{\textbf{M2 \& M3}} & \multicolumn{3}{c}{\textbf{M2 \& M5}}\\
\hhline{~|---|---|---}
\rowcolor{mydarkgray}
{\textbf{no.}} & {$\bm{\mu}_a$} & {$\bm{\mu}_p$} & {$\bm{\tau}_0$} & {$\bm{\mu}_a$} & {$\bm{\mu}_p$} & {$\bm{\tau}_0$} & {$\bm{\mu}_a$} & {$\bm{\mu}_p$} & {$\bm{\tau}_0$}\\
\hline
\rowcolor{mywhite}
Fluid 1 & 9.50\% & 15.79\% & 17.06\% & 9.40\% & 9.64\% & 4.81\% & 9.40\% & 9.64\% & 28.57\%\\
\rowcolor{mylightgray}
Fluid 2 & 1.36\% & 4.19\%  & 16.84\% & 1.54\% & 4.37\% & 16.67\% & 49.23\% & 73.07\% & 25.67\%\\
\rowcolor{mywhite}
Fluid 3 & 2.48\% & 0.36\% & 14.39\% & 5.64\% & 2.87\% & 17.21\% & 50.20\% & 53.96\% & 57.72\%\\
\rowcolor{mylightgray}
Fluid 4 & 2.42\% & 14.33\% & 33.24\% & 10.21\% & 0.17\% & 29.92\% & 10.21\% & 0.17\% & 47.09\%\\
\rowcolor{mywhite}
Fluid 5 & 4.48\% & 2.83\% & 10.87\% & 11.82\% & 10.25\% & 17.83\% & 55.88\% & 57.50\% & 64.35\%\\
\rowcolor{mylightgray}
Fluid 6 & 2.69\% & 0.57\% & 20.73\% & 8.97\% & 6.95\% & 25.92\% & 54.48\% & 55.90\% & 68.03\%\\
\rowcolor{mywhite}
Fluid 7 & 1.47\% & 12.60\% & 32.08\% & 7.20\% & 12.80\% & 17.67\% & 7.20\% & 12.80\% & 38.25\%\\
\rowcolor{mylightgray}
Fluid 8 & 3.11\% & 7.60\% & 25.49\% & 8.36\% & 1.73\% & 29.57\% & 54.18\% & 51.80\% & 69.65\%\\
\rowcolor{mywhite}
Fluid 9 & 5.76\% & 1.03\% & 22.94\% & 10.90\% & 4.48\% & 27.13\% & 55.43\% & 54.76\% & 68.52\%\\
\rowcolor{mylightgray}
Fluid 10 & 9.96\% & 0.06\% & 38.70\% & 10.52\% & 5.83\% & 30.07\% & 10.52\% & 5.83\% & 47.52\%\\
\rowcolor{mywhite}
Fluid 11 & 5.58\% & 0.23\% & 22.38\% & 7.13\% & 1.43\% & 23.71\% & 53.55\% & 53.28\% & 67.14\%\\
\rowcolor{mylightgray}
Fluid 12 & 2.63\% & 0.59\% & 13.36\% & 6.45\% & 4.52\% & 16.70\% & 53.23\% & 54.76\% & 64.13\%\\
\hline
\rowcolor{mylightblue}
{\textbf{\textit{Average}}} & {\textbf{\textit{4.29\%}}} & {\textbf{\textit{5.01\%}}} & {\textbf{\textit{22.34\%}}} & {\textbf{\textit{8.18\%}}} & {\textbf{\textit{5.42\%}}} & {\textbf{\textit{21.43\%}}} & {\textbf{\textit{38.62\%}}} & {\textbf{\textit{40.29\%}}} & {\textbf{\textit{53.89\%}}}\\
\hline
\end{tabular}
\caption{The second column, labelled as ``M2 \& M1'', showcases the systematic error of our mathematical framework (M1) when evaluating the apparent viscosity $\mu_a$, plastic viscosity $\mu_p$, and yield point $\tau_0$, of the non-Newtonian drilling fluids listed in Table \ref{table:non-newtonian-drilling-fluid-data}, with the reference being the mathematical framework of the Fann-35 viscometer or \cite{lam2014interpretation} (M2). Analogously, the subsequent columns, labelled as ``M2 \& M3'' and ``M2 \& M5'', display the corresponding systematic errors associated with the mathematical frameworks of \cite{li2020rheological} (M3) and \cite{sedaghat2017novel} (M5), respectively. The bottommost row presents the average systematic error associated with each (non-reference) mathematical model.}
\label{table:non-newton-fluid-comparison}
\end{table}

As evident from the aforementioned table, the systematic error in plastic viscosity associated with M1 ranges from 0.06\% (Fluid 10) to 15.79\% (Fluid 1), with an average of 5.01\%. Conversely, the systematic error in plastic viscosity associated with M3 varies between the lowest value of 0.17\% (Fluid 4) and the highest value of 12.80\% (Fluid 7), with an average of 5.42\%. While the highest systematic error of M1 is marginally higher than that of M3, the average systematic error of M1 is lower. Moreover, for most of the non-Newtonian drilling fluids, the systematic error in plastic viscosity associated with M1 is significantly lower than that of M3 and is considerably lower than that of M5 in all cases.

Upon closer examination, it is evident that the average systematic error in plastic viscosity associated with M1 for non-weighted non-Newtonian drilling fluids is 10.70\%, slightly higher than that of M3, which stands at 7.11\%. However, in contrast, for weighted non-Newtonian drilling fluids, the average systematic error in plastic viscosity associated with M1 is 2.17\%, slightly lower than that of M3, which amounts to 4.58\%.  

Based on the comprehensive comparative analyses presented above, it can be concluded that our proposed mathematical model exhibits slightly higher accuracy in determining the plastic viscosity of most non-Newtonian drilling fluids compared to the mathematical model developed in \cite{li2020rheological}. Moreover, our model significantly outperforms the mathematical model introduced by \cite{sedaghat2017novel} in all cases. However, when exclusively dealing with non-weighted non-Newtonian drilling fluids, the mathematical model of \cite{li2020rheological} demonstrates a slight advantage over our proposed model. Conversely, when dealing solely with weighted non-Newtonian drilling fluids, our model outperforms the one proposed by \cite{li2020rheological}.

Consequently, both our model and that of \cite{li2020rheological} can be considered viable options for determining the plastic viscosity of various non-Newtonian drilling fluids in the industry. Nevertheless, when assessing the plastic viscosity of a set of non-Newtonian drilling fluids, categorised into weighted and non-weighted types, engineers and scientists might prefer using our model for solely weighted non-Newtonian drilling fluids, while leaning towards the model proposed by \cite{li2020rheological} for solely non-weighted non-Newtonian drilling fluids.

Lastly, we conduct an in-depth analysis of the systematic error data in calculating the yield point of the non-Newtonian fluids listed in Table \ref{table:non-newtonian-drilling-fluid-data} using our model (M1) and compare them with the corresponding systematic error data of the models proposed by \cite{li2020rheological} (M3) and \cite{sedaghat2017novel} (M5). The systematic error associated with M1, in this case, ranges between 10.87\% (Fluid 5) and 38.70\% (Fluid 10), with an average of 22.34\%. Conversely, the systematic error of M3 varies between 4.81\% (Fluid 1) and 30.07\% (Fluid 10) with an average of 21.43\%, while the systematic error of M5 ranges from 25.67\% (Fluid 2) to 69.65\% (Fluid 8) with an average of 53.89\%. These findings reveal that the lowest, highest, and average values of the systematic errors associated with M1 are significantly lower than the corresponding values of M5. However, it is noteworthy that the lowest and highest values of the systematic errors associated with M3 are substantially lower than their corresponding values of M1, and the average systematic error of M3 is slightly lower than that of M1.

Further investigation into the data reveals that the average systematic error of M1, when determining the yield point of weighted non-Newtonian drilling fluids, is 18.38\%, slightly lower than that of M3, which stands at 21.84\%. In contrast, the average systematic error of M1 in determining the yield point of non-weighted non-Newtonian drilling fluids is 30.27\%, considerably higher than that of M3, which is only 20.62\%.

Based on the above comparative analyses, it can be inferred that our proposed model exhibits slightly better accuracy than the one proposed by \cite{li2020rheological} when determining the yield point of solely weighted non-Newtonian drilling fluids, which constitute the majority of the fluids listed in Table \ref{table:non-newtonian-drilling-fluid-data}. However, when dealing with solely non-weighted non-Newtonian drilling fluids, the model proposed by \cite{li2020rheological} demonstrates significantly better accuracy compared to our proposed model. Nevertheless, in a general context, our model performs only slightly worse than that of \cite{li2020rheological} and significantly outperforms the model proposed by \cite{sedaghat2017novel} in all cases.

Consequently, for both industrial and laboratory applications, engineers and scientists may find it beneficial to utilise our mathematical model as a better substitute for the one proposed by \cite{li2020rheological} when determining the yield point of solely weighted non-Newtonian drilling fluids. Nevertheless, we strongly advise against using our model for determining the yield point of solely non-weighted non-Newtonian drilling fluids and suggest employing the model presented by \cite{li2020rheological} for this specific purpose.

In summary, for applications primarily within the drilling and petroleum industry, employing the Marsh funnel in combination with our proposed mathematical model offers a reliable and convenient method to determine the apparent viscosity, plastic viscosity, and yield point of weighted non-Newtonian drilling fluids with average systematic errors of 3.52\%, 2.17\%, and 18.38\%, respectively. Additionally, it can be utilised to determine the apparent viscosity of non-weighted non-Newtonian drilling fluids with an average systematic error of 5.84\%. In these aspects, our proposed model stands as the most accurate one developed to date. However, when determining the plastic viscosity and yield point of non-weighted non-Newtonian drilling fluids, our model yields average systematic errors of 10.70\% and 30.27\%, respectively. In such cases, we recommend adopting the model developed by \cite{li2020rheological} to achieve higher accuracy.

\section{Conclusions and future work}
In conclusion, our current mathematical framework demonstrates potential for substantial improvement in accurately determining the rheological properties of both Newtonian and non-Newtonian drilling fluids. To achieve this, we propose several avenues for advancement. Firstly, the generalisation of the strict assumptions outlined in Section \ref{sec: assumptions} and the development of a more precise expression for the exit velocity, $v_e$, in terms of fluid height, $h$, and flow factor, $f$, compared to the current Eqn. \eqref{exvelmod}, can greatly enhance the accuracy of rheological property determination in both categories.

Turning specifically to the realm of non-Newtonian drilling fluids, selecting a more accurate factor in place of $\frac{1}{2}\left(\frac{h}{H_T}\right)$ and $\frac{1}{4\sqrt{2}}\left(\frac{h}{H_T}\right)$ in the expressions for the average wall shear rate, $\langle \dot{\gamma}_w\rangle(h)$, of non-weighted and weighted non-Newtonian drilling fluids, respectively, as given in Eqns. \eqref{non-weight-non-newton-shear-rate-1} and \eqref{weight-non-newton-shear-rate-1}, will further enhance the model's accuracy in determining the apparent viscosity, plastic viscosity, and yield point of both these categories of fluids.

Furthermore, the consideration of a time-dependent flow factor, rather than a constant value, introduces a more realistic representation of non-Newtonian behaviour, thus enhancing the model's overall accuracy. Lastly, the formulation of precise explicit expressions for wall shear stress, $\tau_w(x,y,z,h)$, and wall shear rate, $\dot{\gamma}_w(x,y,z,h)$, for all points $(x,y,z)\in S(h)$ and all $h\in [0,H_M]$, and incorporating these expressions into Eqns. \eqref{tauw1} and \eqref{gammaw1}, offers the potential for accurate quantification of average wall shear stress, $\langle \tau_w \rangle(h)$, and average wall shear rate, $\langle\dot{\gamma}_w\rangle(h)$, significantly enhancing the determination of rheological properties for both weighted and non-weighted non-Newtonian drilling fluids.

These proposed advancements have the capacity to greatly elevate the accuracy and capabilities of our mathematical model, rendering it an invaluable tool in accurately determining the complex rheological properties of drilling fluids. This, in turn, has the potential to provide substantial benefits to the drilling industry and allied sectors, driving advancements and improvements in various related fields.


\cleardoublepage 
\phantomsection  
\renewcommand*{\bibname}{References}

\addcontentsline{toc}{chapter}{\textbf{References}}

\bibliography{uw-ethesis}

\nocite{*}




\end{document}